\documentclass[11pt,fleqn,a4paper]{article}
\usepackage{titling}
\usepackage{preliminary}
\usepackage{hyphenat}
\usepackage[authoryear]{natbib}
\usepackage{amssymb}
\usepackage{amsthm}
\usepackage{epstopdf}
\usepackage{amsmath}

\usepackage{scrlayer-scrpage}
\usepackage{subfigure}
\usepackage{setspace}
\usepackage[bottom]{footmisc}
\usepackage{endnotes}
\usepackage{graphicx}
\usepackage{textcomp}
\usepackage{natbib}
\usepackage[ansinew]{inputenc}
\usepackage{verbatim}
\usepackage{lscape}
\usepackage{array}
\usepackage{fltpoint}
\usepackage{dcolumn}
\usepackage{booktabs}
\usepackage{rotating}
\usepackage[norounding,point]{rccol}
\usepackage{hyperref}
\usepackage{placeins}
\usepackage{multirow}
\usepackage{paralist}
\usepackage{enumitem}
\usepackage{textgreek}
\usepackage{float}
\usepackage{graphicx}
\usepackage{tablefootnote}
\usepackage{eurosym}

\usepackage{multirow}

\usepackage{tikz-cd}
\usepackage{ctable}

\usepackage{tikz}
\usetikzlibrary{positioning}
\usetikzlibrary{snakes}

\usepackage{multicol}
\usepackage{caption}

\usepackage{subfigure}

\usepackage{amsmath}

\usepackage{geometry}
\numberwithin{equation}{section}

\hypersetup{
colorlinks=true,
urlcolor=blue,
linkcolor=red,
citecolor=blue
}
\defcitealias{AUSSENHANDELSSTATISTIK2018}{AS, 2018a}
\defcitealias{Volksw}{AS, 2018d}
\defcitealias{AFS2020}{AS, 2020}
\defcitealias{Beschaftigungsstatistik2019}{AS, 2019}
\defcitealias{Beschaftigungsstatistik2018}{AS, 2018b}
\defcitealias{Lohnstatistik2018}{AS, 2018c}
\defcitealias{BFS2020}{BFS, 2020}

\newcommand*{\captionsource}[2]{%
  \caption[{#1}]{%
    #1%
    \\\hspace{\linewidth}%
    \textbf{Source:} #2%
  }%
}

\setheadwidth{textwithmarginpar}
\usepackage{setspace}
\usepackage{xspace}

\numberwithin{theorem}{section}
\numberwithin{assumption}{section}

\newcommand{\auslassen}[1]{}
\newcommand{\bb}[1]{\textcolor{blue}{[bb: #1]}}
\newcommand{\sg}[1]{\textcolor{Green}{[sg: #1]}}

\renewcommand{\bb}[1]{}
\renewcommand{\sg}[1]{}
\usepackage{graphicx}

\begin{document}
\begin{center}
{\LARGE How residence permits affect the labor market attachment of foreign workers: Evidence from a migration lottery in Liechtenstein}

\large \vspace{0.8cm}

{\large Berno Buechel$^\dagger$, Selina Gangl$^\dagger$, Martin Huber$^{\dagger *}$}\medskip

{\small {$^\dagger$University of Fribourg, Department of Economics} }\\
{\small {$^*$Center for Econometrics and Business Analytics, St.\ Petersburg State University} }

\today

\end{center}

\begin{abstract}
We analyze the impact of obtaining a residence permit on foreign workers' labor market and residential attachment. To overcome the usually severe selection issues, we exploit a unique migration lottery that randomly assigns access to otherwise restricted residence permits in Liechtenstein (situated between Austria and Switzerland). Using an instrumental variable approach, our results show that lottery compliers (whose migration behavior complies with the assignment in their first lottery) raise their employment probability in Liechtenstein by on average 24 percentage points across outcome periods (2008 to 2018) as a result of receiving a permit. Relatedly, their activity level and employment duration in Liechtenstein increase by on average 20 percentage points and 1.15 years,  respectively, over the outcome window. These substantial and statistically significant effects are mainly driven by individuals not (yet) working in Liechtenstein prior to the lottery rather than by previous cross-border commuters. 
Moreover, we find both the labor market and residential effects to be persistent even several years after the lottery with no sign of fading out. These results suggest that granting resident permits to foreign workers can be effective to foster labor supply even beyond the effect of cross-border commuting from adjacent regions.
\end{abstract}

{\small  \textbf{Keywords:} international migration, cross-border commuting, natural experiment, lottery}

{\small  \textbf{JEL classification:} F22, J61.  \quad }

\begin{singlespace}
\smallskip {\scriptsize We are grateful to the Government of Liechtenstein for the permission to realize this research project and to the Immigration and Passport Office as well as the Office of Statistics of Liechtenstein for 
their valuable support concerning data provision and processing. We have benefited from comments by  Andreas Brunhart and Andreas Steinmayr. 
 Addresses for correspondence:  Berno Buechel, Selina Gangl,  and Martin Huber, University of Fribourg, Bd.\ de P\'{e}rolles 90, 1700 Fribourg, Switzerland; berno.buechel@unifr.ch, selina.gangl@unifr.ch, martin.huber@unifr.ch.}
 \end{singlespace}

\thispagestyle{empty}\pagebreak

\newgeometry{top=3cm,bottom=3cm,right=2.5cm,left=2.5cm}

{\small \renewcommand{\thefootnote}{\arabic{footnote}} %
\setcounter{footnote}{0}  \pagebreak \setcounter{footnote}{0} \pagebreak %
\setcounter{page}{1} }


\section{Introduction}\label{s:intro}

Most labor markets rely on both locally-born and foreign-born workers. For instance, the share of foreign-born persons in the U.S.\ labor market corresponds to 17.4\% in 2019, amounting to 28.4 million people.\footnote{Reported by U.S.\ Bureau of Labor Statistics, https://www.bls.gov/news.release/forbrn.nr0.htm/labor-force-characteristics-of-foreign-born-workers-summary, retrieved 2021-02-11.} World-wide, the number of migrant workers is estimated to be 164 million in 2017, with 23\% in North America, 32\% in Europe, and 13\% Asia.\footnote{Reported by International Labour Organization, https://migrationdataportal.org/themes/labour-migration, retrieved 2021-02-11.} While migration from low to higher income countries receives a lot of attention in public and scientific discussions, an immense amount of labor mobility is realized between rather developed nations, likely competing for skilled workers.
To address the high demand for foreign labor, offering residence permits for migrant workers may seem a rather obvious policy. 
In many regions and countries, however, alternative policies such as admitting seasonal workers or cross-border commuters play an important role. 
In France, for instance, about 438k workers commute to another country every day for work.\footnote{Reported by EUROSTAT, https://ec.europa.eu/eurostat/statistics-explained/index.php?title=Archive:Statistics\_on\_commuting\_patterns\_at\_regional\_level\&oldid=463740, retrieved 2021-02-11.} \bb{better incoming not outgoing}
An interesting question that arises in this context is 
whether residence permits increase foreign labor market attachment, relative to alternatives such as cross-border commuting. This may be beneficial in an environment of increased skill competition in order to attract and retain key workers required for fostering economic growth. \bb{Eine wichtige Frage fuer die Interpretation ist, was die bisherigen Nicht-Grenzgänger tun, wenn sie die Lotterie verlieren: 1.\ Commute, 2.\ Arbeit nicht antreten.} \sg{To do: Analyse der Verlierer, die nicht Grenzgänger waren. Reicht eine Statistik aus? Wie viele Verlierer und Nicht-Grenzgänger pendeln. Nur nach 1. Lotterie-Teilnahme???}

In this paper, we assess the causal effect of obtaining a residence permit 
 on the labor market and residential attachment of foreign workers based on an annual migration lottery that is unique in Europe. In Liechtenstein, a wealthy microstate that is situated between Austria and Switzerland, 
 two lottery draws for residence permits are held every year. The lottery randomly assigns access to residence permits among applicants from the European Economic Area (EEA)\footnote{Iceland, Liechtenstein, and Norway plus all EU member states (Austria, Belgium, Bulgaria, Croatia, Cyprus, Czech Republic, Denmark, Estonia, Finland, France, Germany, Greece, Hungary, Ireland, Italy, Latvia, Lithuania, Luxembourg, Malta, Netherlands, Poland, Portugal, Romania, Slovakia, Slovenia, Spain, and Sweden) are part of this free trade and movement agreement.} who hold an employment contract with a company in Liechtenstein.\footnote{For EEA citizens who do not belong to the working population and can finance their livelihood from their own resources, there is another lottery that is conducted at the same time.} We are the first to link European migration lottery data with administrative data on individual labor market records. We exploit the random assignment as instrument for the reception of a residence permit that is conditional on actually moving to Liechtenstein. This allows us to assess the local average treatment effect (LATE) of moving among compliers of the first lottery they participated in, i.e.\ those whose migration behavior complies with the assignment obtained in their first lottery participation, making up 36\% of our sample. We apply a flexible instrumental variable (IV) estimator based on inverse probability weighting (IPW), in which we reweigh the outcomes by the inverse of the conditional instrument probability given the lottery year, the so-called instrument propensity score. This enables us to  control for the fact that the share of lottery winners changes over the years as a function of lottery applicants, which is important in order to avoid confounding, e.g.\ due to business cycle effects.

We find that receiving a residence permit statistically significantly increases the employment probability by 24 percentage points on average over our outcome periods 2008 to 2018. Likewise, it increases the activity level by 20 percentage points, and the employment duration in Liechtenstein by 1.15 years. 
These substantial labor attachment effects remain robust to the inclusion of further covariates in the IPW estimator like gender, age, and nationality. Assessing effect heterogeneity across previous cross-border commuters and new labor market entrants who (despite holding a work contract) have not worked in Liechtenstein yet when participating in the lottery, the effects are stronger and statistically significant in the latter group. 
This suggests that residents permits are more effective for attracting new labor market entrants than for keeping previously commuting workers.

Moreover, we assess the effect of receiving the residence permit on residence in Liechtenstein two and more years after the lottery. 
  We find the residence probability and duration to significantly increase by 71 percentage points and 3.44 years, respectively, on average, with little differences between previous cross-border commuters and new labor market entrants. Finally, we consider the labor market and residential effects separately for specific years after the lottery and find them to be persistent with no sign of fading out. That is, even in the tenth year after the first lottery participation, the impact on the employment and residence probability is comparable to the respective average effect across all periods and statistically significant at the 5\% level, albeit confidence intervals are large due to the smaller sample size. We argue that tax advantages when moving to Liechtenstein are likely a very important (if not the main) mechanism for driving our strong effects, as confirmed by a survey among foreign workers in Liechtenstein. \bb{diskussion ist noch unausgereift, aber ok für den Moment.}

Very broadly speaking, our paper fits into the literature on labor migration, see for instance \cite{FASANI2020}. More specifically, it is among a relatively scarce number of studies that exploit migration lotteries to convincingly assess the causal effect of residence permits on labor market behavior or related outcomes.  \citet{Gibson2011}, for instance, investigate a lottery in the Pacific island state of Tonga for residence permits in New Zealand and study welfare effects on household members of families left behind.   \cite{Gibson2017} use the same lottery and find positive income effects among migrants themselves. \cite{Clemens12} analyze a lottery in a specific multinational firm that allocated U.S.\ visas to Indian software workers and conclude that migration to the U.S.\ entails a sixfold increase in wages.
\cite{MERGO2016} considers the U.S.\ Diversity Visa lottery for Ethopians and finds that their migration to the U.S.\ increases welfare (in particular consumer expenditure) of the families left behind in Ethiopia. 
\cite{Mobarak2020} examine a visa lottery for low-skilled workers from Bangladesh intending to work in the palm-oil industry in Malaysia; and find that migration leads not only to a substantial income rise among migrants, but also to an increase in the household consumption of the family left behind.

While the previously mentioned studies consider migration from a less developed to a more developed country, 
 a rather unique feature of our lottery is that it exclusively concerns member states of the European Economic Area (EEA). Therefore, our paper contributes to the literature by considering labor migration between rather developed and wealthy countries. This is important, because 
a large amount of labor mobility is realized between rather developed nations, which are likely competing for skilled labor. A second important distinction is that we focus on the labor force attachment of individuals that could resume or start working in Liechtenstein even without living there, i.e.\ by means of cross-border commuting, in particular from nearby Austria or Switzerland, which is actually the most common form of labor in Liechtenstein.\footnote{Indeed, about three thirds (76.9\%) of the 29k 
 foreigners who work in Liechtenstein commute cross-border \citep{Beschaftigungsstatistik2018}.} Our paper therefore sheds light on whether residence permits (and the associated amenities like tax reductions) incentivize foreigners to remain in the labor market, which appears an important piece of information for policy makers in a competitive open economy with a high demand for foreign labor, as it is the case in Liechtenstein.\footnote{See \citet{HuberNowotny2013} for an empirical study on which personal characteristics drive the willingness to commute and migrate across borders in regions of the Czech Republic, Hungary, and Slovakia that are situated close to Austria. The intention to commute or migrate is for instance found to be significantly negatively associated with age and being a female, and significantly positively associated with being single or feeling deprived when comparing the own social status to peers. Also higher education has a positive correlation, which is, however, not significant at the 5\% level.  Therefore, personal factors most likely play a role for the question which type of workers respond to specific incentives like cross-border work permits or residence permits, even though it needs to be pointed out that the findings in \citet{HuberNowotny2013} do not necessarily directly carry over to the context of Liechtenstein.}  As a motivation for the potential economic relevance of labor migration regulations, see for instance \cite{Beerli}. The authors find for Switzerland that reducing restrictions for cross-border workers increased the size and productivity of skill-intensive sectors (in particular those with previous skill shortages) such that even the wages of highly educated natives rose, despite the hike in foreign employment. 
  Our paper appears to be the first that relates the literature on migration lotteries with the topic of commuting, in particular, cross-border commuting.

\bb{very short survey of commuting literature and our contribution: Well, I did not find nice papers about cross-border commuting.}

The remainder of the paper is structured as follows. Section~\ref{s:background} provides information about Liechtenstein and its migration lottery. Section~\ref{s:data} introduces our data and provides descriptive statistics. Section~\ref{s:metrics} discusses the empirical strategy.  Section~\ref{s:results} presents and interprets the results. Finally, Section~\ref{s:concl} concludes.

\section{Institutional background} 
\label{s:background}

\begin{figure}[h!]
\begin{center}
{ \includegraphics[scale=.3]{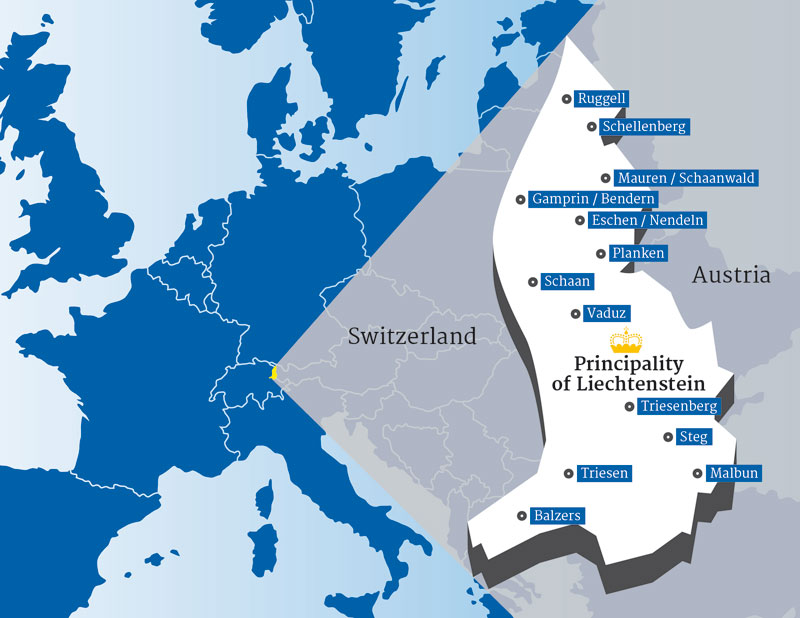}}
\caption{Map of Liechtenstein (Source: Liechtenstein Marketing)}
\label{fig:map}
 \end{center}
 \end{figure}

This section gives a very brief overview of the economy, the labor market, and  the migration lottery of Liechtenstein.\footnote{More details about the institutional background are provided in Appendix~\ref{s:app_background}.}
As illustrated in Figure~\ref{fig:map}, Liechtenstein is a micro state situated in Central Europe, between Switzerland in the West and Austria in the East. The official language is German. Liechtenstein's population amounts to almost 40k inhabitants, while its labor force is of roughly the same size -- in fact,  slightly exceeding the population.  
Liechtenstein is a small open economy. Since 1923 it has a customs union with Switzerland 
 and since 1995 it is member of the European Economic Area (EEA), which includes all European Union (EU) states plus Norway and Iceland, but not Switzerland.
 Export of goods, excluding trade with and via Switzerland, account for 55\% of its GDP. Its most important industries are mechanical engineering and provision of financial and insurance services, 
 which account for 16.2\% and 13.3\% of the GDP, 
  respectively \citep{Volksw}. Liechtenstein is among the wealthiest countries in the world with a nominal GDP per employed person of  about 200k USD. 
The official currency in Liechtenstein is the Swiss Franc (CHF), which had an average exchange rate of 1.04 USD/CHF in the last decade.

 The labor market in Liechtenstein is characterized by a low unemployment rate  and a large demand for foreign labor. The strong economic growth in recent decades in combination with the small size of the country fueled an ongoing employment expansion. Table \ref{tab: Number of employees in Liechtenstein} documents the increase in the number of employees from 1980 onwards and also distinguishes between employees residing in Liechtenstein and cross-border commuters, which have grown even more substantially than the total labor force. Since 2010  there are more cross-border commuters in the work force than employees residing in Liechtenstein. Most employees work in the service sector (61.9\%) followed by the industrial sector (37.4\%). 

  Wages in Liechtenstein are relatively high when compared to other Western European countries, which is most likely an important pull factor for attracting foreign labor. The median gross wage per month is about 7k USD, 
   which is similar to 
    the level of neighboring Switzerland, and substantially higher than in most EU countries, including neighboring Austria. 
   The gross median income of cross-border commuters 
   (CHF 6'723) is similar to that of residents (CHF 6'612) \citep{Lohnstatistik2018}.
Cross-border commuters are predominantly male (64.4\%), working in the tertiary sector (55\%), living in Switzerland or Austria (96\%), and holders of a citizenship of a member country of the EEA (62.2\%) \citepalias{Beschaftigungsstatistik2018}.

 \begin{table}[h!]
 \caption{Number of employees in Liechtenstein}
\label{tab: Number of employees in Liechtenstein}
\begin{center}

\begin{tabular}{|l|c|cc|}
\hline\hline
 &   \multicolumn{3}{c|}{Employees in Liechtenstein}\\
 \hline
 year & Residing in Liechtenstein & Cross-border commuters & Total\\
 \hline
1980  & 11,543 & 3,297 &14,840\\
1990 & 13,020 & 6,885 & 19,905\\
2000 & 15,605 & 11,192 & 26,797\\
2010 & 16,764 & 17,570 & 34,334\\
2017 & 17,362 &21,299 & 38,661\\
2018 & 17,597 & 22,038 &39,635\\
\hline
\end{tabular}
    \par
{\footnotesize Source: \citet{AFS2020}; ``residing in Liechtenstein'' is self-calculated}

\end{center}

\end{table}

One important reason for the high share of cross-border commuters among the labor force is regulated access to residence permits in Liechtenstein. Despite being a member of the EEA, Liechtenstein is permitted to restrict residence of EEA citizens in Liechtenstein.
However, by the EEA treaty, Liechtenstein is required to issue at least 56 residence permits for the purpose of employment every year, half of which must be assigned by a lottery.\footnote{Despite the restrictive rules for immigration to Liechtenstein there is an inflow of 17.0 (net inflow of 4.3) immigrants per 1,000 inhabitants \citep{Migrationsstatistik2019}. The dominant formal purpose for immigration to Liechtenstein is family reunification. The fraction of foreigners among the residents in Liechtenstein is 34\% \citep{AFS2020}.}


Holding at least one lottery per year is required by law (see Law on the Free Movement of Persons for EEA and Swiss nationals (2009), section 39, 1). Usually, two lotteries take place per year, one in spring and one in fall \citep{APA2020}. Each lottery consists of two stages, namely the pre-draw and the final draw  \citep[section 37, 2]{PFZG2009}. Participants must win in both parts of the lottery to receive a residence permit. Requirements for participation include holding an EEA citizenship and paying the participation fees in time.  In the final draw, participants must also provide an employment contract of more than one year with a minimum activity level of 80\% or, else, an authorized permanent cross-border business activity in case of self-employment \citep{APA2019}.
After winning both the pre-draw and the final draw, the lottery participant has six months time to relocate to Liechtenstein, otherwise the residence permit expires \citep[section 37, 2]{PFZG2009}. For this reason, our treatment is defined based on residing in Liechtenstein in the year after the lottery, as obtaining the permit is tied to actually moving there. The drawing of winners is done blindly by hand. This procedure is monitored by at least one judge. \bb{wer genau.} \sg{ein Landrichter.}

Lottery losers of either stage may participate again in subsequent lotteries, while multiple applications to the very same lottery are not allowed \citep[section 38, 1) c)]{PFZG2009}. As the decision to repeatedly take part in the lottery is most likely
 endogenous to the first lottery outcome, our main evaluation strategy relies on the first lottery participation of an individual in our data window. \bb{it is certainly dependent on the outcome of the first lottery. Isn't the point rather that there could be heterogeneity in how often losers would try, inducing a selection issue, as I argue further below?} Furthermore, as participation in the final draw is conditional on succeeding in the pre-draw only, we base our instrumental variable approach on the pre-draw alone.

The incentives to participate in the lottery are arguably related with the costs and benefits of residing in Liechtenstein. For most lottery participants, the relevant alternative is to reside in a neighboring country and commute cross-border.\footnote{For other participants it can be the case that they will stop working in Liechtenstein or not start working in Liechtenstein, despite holding an employment contract.} In a  survey of cross-country commuters to Liechtenstein \citet[p.\ 57]{Marxer2016} ask about the reasons to move to Liechtenstein, given the presumption that the respondents would move there in the future. The top answer is ``taxes and duties'' (86\% of respondents), which is even ticked more often than ``proximity to the workplace'' (80\% of respondents), while all other categories are ticked by less than 22\% of respondents. Indeed, taxes are substantially lower in Liechtenstein than in Switzerland such that the net disposable income of given gross incomes and given household types is about 10 percentage points higher in Liechtenstein \citep{Brunhart2016}. In Austria taxes are even substantially higher than in Switzerland such that the net disposable income there is likely even lower, despite the lower costs of living. 
Accordingly, \cite{Marxer2016} find that 31\% of the cross-border commuters living in Switzerland and 75\% of those living in Austria are not satisfied with their tax system; 
while there is a is a particularly low willingness to move to Liechtenstein for those employees who pay taxes in Liechtenstein (these are employees who reside in Austria and work in Liechtenstein's public sector).

The second motivation to participate in the lottery, living closer to the workplace, appears obvious,  but must be put in perspective: most cross-border commuters have quite short commutes. 59\% of them travel less than 30 minutes to work and only 6\% more than 1 hour \citep[p.\ 36]{Marxer2016}. Still, residing in Liechtenstein can lead to more convenient commutes, be it because of even shorter commuting times, fewer bus or train changes when using public transport, or different means of transport (e.g. biking instead of driving). 
Given these advantages of residing in Liechtenstein over commuting to Liechtenstein, we note that 
 financial incentives are probably the most important factor for participating in the lottery, while distance to the workplace also matters.

%

\section{Data}\label{s:data}

This section provides a description of our data set and the key variables  along with descriptive statistics. 
Our data base was created by linking records from the migration lottery with employment statistics in Liechtenstein. The lottery records cover all lottery participants from 2003 to 2019. In particular, they include information on when and how often an individual applied to the migration lottery. This enables us to define the instrument based on whether an applicant has won the pre-draw or not in the first lottery participation. In addition, the data contain personal characteristics such as the year of birth, nationality, and gender, which are asked in the application form for the lottery. 

The employment statistics cover the years 2005 to 2018. Every employer in Liechtenstein is obliged to report new employment entries and exits on a monthly base. At the end of each year, companies receive a list of their reported employees for proofreading and are obliged to resubmit a corrected version  \citep{Beschaftigungsstatistik2019}. The employment statistics contain variables characterizing the labor market behavior of the applicants. This includes information on whether an individual has worked in Liechtenstein in the year prior to lottery participation and whether she or he has started or continued dependent or self-employment in the years after  lottery participation. For each year, also the activity level in percent is reported, as well as the country of residence. Finally, several personal characteristics are observed that are also available in the migration lottery records, namely the year of birth, nationality, and gender. Whenever there are differences in these variables across the two data sources, we prioritize the employment statistics which we suspect to be of higher quality, as they are repeatedly provided and checked. In contrast, the lottery records only contain information that was originally handwritten in the application form. 
 Linking both data sets is based on a unique personal identifier and the created data base is fully anonymized.

\bb{hier Paragraph zu SVERWEIS gestrichen.}

\begin{figure}[h!]
\begin{center}
{ \includegraphics[scale=.65]{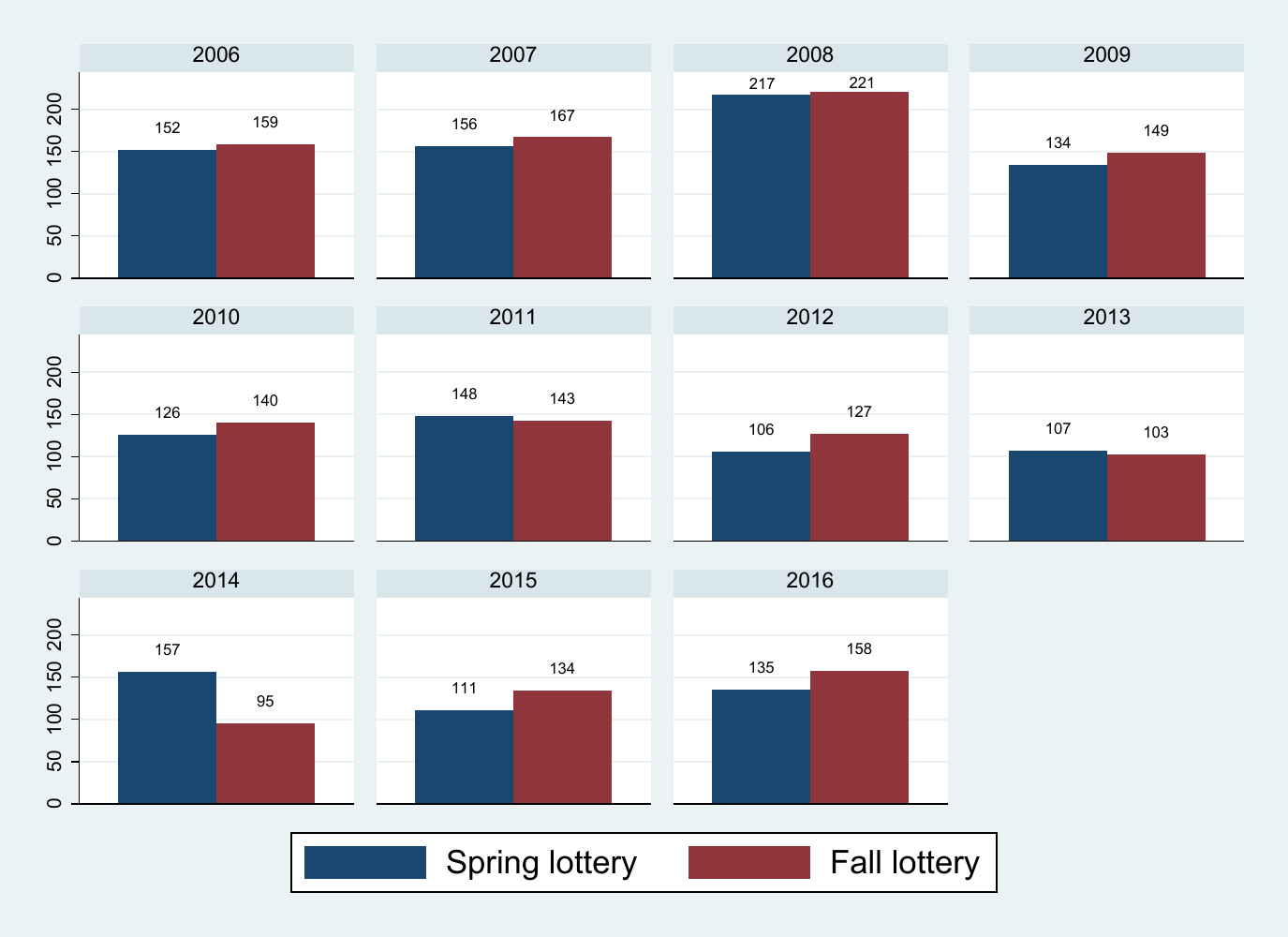}}
\caption{\label{figpart} Annual number of first lottery participations}
 \end{center}
 \end{figure}

In total, the migration lottery data contain 9,906 observations from 2003 to 2019. While each lottery draw is random, the possibility to repeatedly participate in case of losing might induce a selection problem, as more persistent applicants who participate more than once in the lottery likely differ in terms of their characteristics from the initial pool of applicants. 
We overcome this concern by exclusively considering the first lottery participation in our data window, which reduces the sample to 5,091 observations. Hence, we compare individuals who won when first participating in the lottery with those who lost, but might  have participated again and won in a later lottery. This strategy yields conservative effects in the sense that they likely provide a lower bound to those of a hypothetical comparison of winning vs.\ losing and being prevented from any further lottery participation.  \bb{This is a conservative strategy. If correct, we can make this explicit.} Since the employment statistics are only available from 2005 onwards, we restrict the sample of first lottery participants to the years 2006 or later, in order to observe the labor market state of each applicant in the year prior to the lottery. This will be important for our analysis of effect heterogeneity across previous cross-border commuters and new labor market entrants. Another sample restriction comes from the fact that the last period in which outcomes are observed in the employment statistics is 2018. This requires us to consider 2016 as last lottery year, because outcomes are measured at the earliest 2 years after the lottery, as it will become clear from the discussion further below. Figure \ref{figpart} shows the annual number of the first lottery participants from 2006 to 2016, separately for the spring and fall lotteries. Moreover, Figure~\ref{figpart} indicates that this number varies across years, which is also true for the number of all lottery participants, with a peak during the financial crisis in 2008.\footnote{The high number of lottery applications may partly be driven by more applications from individuals previously not working in Liechtenstein, aiming to escape the crisis-induced deteriorating labor market conditions in their home country, as suggested by descriptive statistics in Table \ref{descr_non-cross-border} below. Furthermore, the higher number might partly be caused by cross-border commuters suspecting a larger chance of losing employment when being a commuter rather than a resident, in line with findings in \cite{Kuptsch_2012} that migrants face disproportionately higher risks of job loss in case of economic woes.} Thus, the odds of winning change over time, as the amount of lottery-assigned permits is not 
 deterministic in the number of (first) applications. This implies that the lottery year is a likely confounder of our instrument variable assignment, as the year is likely associated with labor market outcomes through the business cycle. We therefore control for lottery year dummies in our IV approach and include the additional control variables age, nationality, and gender in a robustness check. In sum, our evaluation data set contains 3,145 participants, out of which 350 win the pre-draw in their first participation.

\begin{figure}[ht!]
\centering
\begin{tikzpicture}[snake=zigzag, line before snake = 5mm, line after snake = 5mm]
    \draw[ ->]  (-2,0) -- (12,0);

    \foreach \x in {-2, 3,7,10}
      \draw (\x cm,3pt) -- (\x cm,-3pt);

      \draw (-2,0) node[below=3pt] {$ t=-1 $} node[above=3pt] {Time-dependent covariates };
    \draw (3,0) node[below=3pt] {$ t=0 $} node[above=3pt] { First lottery participation  };
        \draw (7,0) node[below=3pt] {$ t=1 $} node[above=5pt] { Treatment  };
    \draw (10,0) node[below=3pt] {$ t\geq 2 $} node[above=3pt] { Outcome periods };

  \end{tikzpicture}
      \caption{Timeline of measured variables}
      \label{figtime}
  \end{figure}
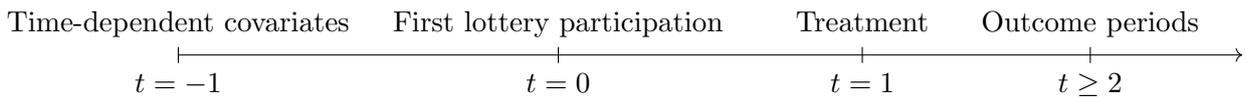

Figure \ref{figtime} provides a timeline for the measurement of the key variables in our analysis, with $t$ denoting a specific year. The instrument, namely the lottery assignment which we henceforth denote by $Z$ (with $Z=1$ for winning and $Z=0$ for losing), is measured in the year of the first lottery participation, which is our baseline period ($t = 0$). The treatment (denoted by $D$), namely whether someone has moved to Liechtenstein ($D=1$) (and is thus in possession of a residence permit) or not ($D=0$), is measured one year later ($t = 1$). \bb{which one: permit or moved?} The outcome periods start two years after the lottery (t $\geq$ 2) and continue until the end of the data window for the respective observation, at most up to 12 years after the lottery for someone participating in 2006 with the final outcome being observed in 2018. All in all, our evaluation sample includes 20,009 outcome observations. Personal characteristics (e.g.\ nationality) are generally measured in the year prior to the first lottery participation ($t = -1$), even though those variables that are independent from or deterministic in time (gender and age) may also be obtained from different periods. In some cases there are differences between the personal characteristics in the migration lottery records (stemming from the application form) and the employment statistics (regularly provided by the employer). As the data quality of the employment statistics appears to be higher than that of the lottery records, priority is given to the former when measuring these characteristics.  We henceforth denote the control variables by $X$, which either only contain period dummies for the lottery years in the main specification, or also additional characteristics in our robustness checks.

To check for violations of the random assignment of residence permits through the lottery, Table \ref{descr} reports descriptive statistics on personal characteristics separately for winners ($Z=1$) and losers ($Z=0$) of the first lottery in our evaluation sample. For either group, the mean and the standard deviation (std.dev) of the variables are reported as well as the mean differences across groups along with t-values and p-values.  Among those with non-missing information in the respective personal characteristics,  age, gender, and the various dummy variables for nationality do not differ importantly or statistically significantly (at any conventional level of significance) across groups, thus pointing to a fair lottery. We also see from the table that the majority of the lottery participants is male, of either Austrian or German nationality, and on average 37 to 38 years old.\footnote{In this context, we note that Swiss nationals are not allowed to participate in the lottery. The reason that their share amounts to 1\% in our data is most likely due to holding a second citizenship from the EEA but reporting the Swiss nationality in the employment statistics.}

In contrast to the observed characteristics, the probability of missing information in nationality and age is statistically significantly different across winners and losers, albeit very small in absolute terms (amounting to only 1 percentage point). This difference is, however, most likely caused by an imbalance in missingness across our two data sources rather than a failure of the lottery.  To see this, note that for any individual already working in Liechtenstein prior to the lottery participation, we have access to information from the employment statistics, in which case there is no missing information. For those not working in Liechtenstein prior to the lottery, we need to rely on the variables from the migration lottery, in which missings do occur. Albeit some of the missing information can be used based on the employment statistics in later (i.e.\ treatment or outcome) periods in particular for determining the age in the year of the lottery, this is dependent on entering the labor market in Liechtenstein at some point in time. As lottery losers enter the labor market less frequently than lottery winners, their share of missing covariates is endogenously higher even under a satisfaction of randomized assignment. \bb{say why this is not a big issue.} This issue does not affect  our main results since we do not drop any observations with missing covariate information, in order to avoid jeopardizing randomization through an endogenously selected subsample.

\begin{table}[!h]
\center
\footnotesize
\caption{Descriptive statistics: First participation from 2006 to 2016}
\label{descr}
\begin{center}
 \resizebox{\linewidth}{!}{
\begin{tabular}{lcc|cc|ccc|c}
 \hline\hline
   & \multicolumn{2}{c|}{$Z=1$} & \multicolumn{2}{c|}{$Z=0$} & & & \\
  \hline
 & mean & std.dev & mean & std.dev & mean difference & t-value & p-value & 
  observations \\
  \hline
Female  & 0.29 & 0.45 & 0.30 & 0.46 & -0.01 & -0.51 & 0.61 & 3,145\\
[0.15cm]
    \textit{Nationality}  & & & & & & & \\
  Missing Dummy & 0.00 & 0.00 & 0.02 & 0.13 & -0.02 & -6.76 & 0.00 & 3,145 \\
  Austria  & 0.38 & 0.49 & 0.37 & 0.48 & 0.01 & 0.40 & 0.69 & 3,100 \\
  Germany & 0.39 & 0.49 & 0.42 & 0.49 & -0.02 & -0.88 & 0.38 & 3,100  \\
  Italy & 0.06 & 0.24 & 0.07 & 0.26 & -0.01 & -0.88 & 0.38 & 3,100  \\
  Switzerland & 0.01 & 0.09 & 0.01 & 0.08 & 0.00 & 0.53 & 0.59 & 3,100  \\
   Others & 0.16 & 0.37 & 0.14 & 0.34 & 0.02 & 1.11 & 0.27 & 3,100  \\
   [0.15cm]
    \textit{Age}  & & & & & &  &\\
 Missing Dummy & 0.01 & 0.09 & 0.02 & 0.15 & -0.01 & -2.52 & 0.01 & 3,145 \\
  Age & 37.25 & 9.25 & 37.49 & 9.62 & -0.24 & -0.46 & 0.65 & 3,078\\
  [0.15cm]
\textit{First lottery participation}  & & & & & & & \\
  Dummy 2006  & 0.09 & 0.29 & 0.10 & 0.30 & -0.01 & -0.31 & 0.76 & 3,145 \\
  Dummy 2007 & 0.09 & 0.29 & 0.10 & 0.30 & -0.01 & -0.57 & 0.57 & 3,145 \\
  Dummy 2008 & 0.09 & 0.29 & 0.14 & 0.35 & -0.05 & -2.97 & 0.00 & 3,145 \\
  Dummy 2009 & 0.10 & 0.30 & 0.09 & 0.28 & 0.01 & 0.84 & 0.40 & 3,145 \\
  Dummy 2010 & 0.06 & 0.24 & 0.09 & 0.28 & -0.02 & -1.74 & 0.08 & 3,145\\
  Dummy 2011 & 0.11 & 0.31 & 0.09 & 0.29 & 0.01 & 0.85 & 0.39 & 3,145\\
  Dummy 2012 & 0.09 & 0.28 & 0.07 & 0.26 & 0.02 & 1.02 & 0.31 & 3,145 \\
  Dummy 2013 & 0.08 & 0.28 & 0.06 & 0.25 & 0.02 & 1.17 & 0.24 & 3,145\\
  Dummy 2014 & 0.10 & 0.30 & 0.08 & 0.27 & 0.03 & 1.50 & 0.13 & 3,145 \\
  Dummy 2015 & 0.09 & 0.28 & 0.08 & 0.27 & 0.01 & 0.75 & 0.45 & 3,145\\
  Dummy 2016 & 0.08 & 0.28 & 0.09 & 0.29 & -0.01 & -0.73 & 0.46 & 3,145 \\
  \hline
   Number of observations & 350 &  & 2,795 &  &  &  &  &  \\
   \hline
\end{tabular}
}\\
 {\footnotesize Sources: Lottery data (2005 - 2016) and employment statistics (2005 - 2016); calculations are done by ourselves\\
  \par}
\end{center}
\par

\end{table}

Table \ref{descr} also reports the year dummies for the first lottery participation across instrument values, providing information about variation in the ratio of pre-draw winners and losers across different years. For year 2008, the mean difference in dummies is statistically significant at the 1\% level, owing to the large number of  lottery applicants in that year (see Figure \ref{figpart}), a likely consequence of the financial crisis of 2007-2008. Such potential business cycle-related confounding motivates controlling for period dummies in our IV approach.

\newpage

\section{Econometric approach}\label{s:metrics}

In this section, we discuss our instrumental variable (IV) approach for evaluating local average treatment effect (LATE), see \cite{Imbens+94} and \cite{Angrist+96}, among lottery compliers, i.e.\ among those who are induced to move to Liechtenstein in the year after the lottery by winning. Following \cite{Abadie00}, we assume that our lottery $Z$ is a valid and relevant instrument conditional on covariates $X$, which either include the lottery year dummies (main specification) or both the year dummies and additional personal characteristics (robustness check). To formally state the identifying assumptions, we make use of the potential outcome notation, see for instance \cite{Rubin74}. We denote by $Y(z,d)$ the potential outcome (e.g.\ hypothetical employment) under specific instrument and treatment states $z,d$ $\in$ $\{1,0\}$, and by $D(z)$ the potential treatment state as a function of the instrument assignment.

Conditional IV validity, as formally stated in equation \eqref{eqn}, consists of two parts: The first part (i) implies that the lottery assignment is as good as random given $X$ and thus not associated with other factors affecting the treatment and/or the outcome. For reasons discussed in Section~\ref{s:data}, this appears plausible conditional on the year dummies. The second part (ii)  states that the lottery assignment must not have a direct effect on the outcome other through the treatment, such that the IV exclusion restriction holds. This assumption is satisfied if winning or losing the lottery does not  directly affect the employment decision conditional on the moving decision. Hence, the assumption excludes, for instance, that winning or losing the lottery induces sufficiently strong feelings of appreciation or disappointment that would make the participant change her labor market status.
\vspace{-.75cm}
\begin{eqnarray}
(i)  \quad  Z  \bot \left(D(1), D(0), Y(1,1), Y(1,0), Y(0,1),  Y(0,0))| X \right. \nonumber \\
(ii)  \quad \Pr(Y(1, d) = Y(0, d) = Y (d)|X)=1 \quad \textrm{for } d  \in \{1,0\}
\label{eqn}
\end{eqnarray}


Equation \eqref{mon} formalizes the conditional monotonicity assumption, which rules out the existence of so-called defiers, i.e.\ of individuals that would move to Liechtenstein in the year after the lottery if losing it, but would not move if winning. Since lottery losers are generally not allowed to move to Liechtenstein,\footnote{Exceptions are that someone gets married or has a common child with a resident of Liechtenstein.} this assumption holds by design in our context.
\vspace{-.75cm}
\begin{equation}
 \Pr (D(0) >  D(1) |X) = 0 \label{mon}
\end{equation}
Equation \eqref{sup} is a common support assumption, implying that for any value of $X$, both lottery winners and losers do exist.  Indeed we find in our data that winners and losers appear in any lottery year and as well across age groups, nationalities and gender.
\vspace{-.75cm}
\begin{equation}
 0 \textless \Pr(Z=1|X)\textless 1\label{sup}
\end{equation}
Finally, equation \eqref{relevance} states that the instrument is relevant in the sense that it affects the treatment decision conditional on $X$. As discussed in Section~\ref{s:results} below, winning the pre-draw does indeed importantly and statistically significantly affect the decision to move to Liechtenstein given the control variables.
\vspace{-.75cm}
\begin{equation}\label{relevance}
 \Pr(D=1|Z=1, X) - \Pr(D=1|Z=0, X) \neq 0
\end{equation}

Under these assumptions, the LATE is nonparametrically identified, see \cite{FROELICH2007}, for instance by reweighing observations based on the inverse of the conditional instrument probability $\Pr(Z=1|X)$, known as the instrument propensity score. Equation \eqref{LATE} presents the identification result based on such an inverse probability weighting (IPW) approach as suggested in \cite{FROELICH2007} and \cite{Tan2006}. It is worth noting that the nominator provides the intention-to-treat effect (ITT) or reduced form effect of the lottery assignment $Z$ on the outcome $Y$, which is a weighted average of the LATE on compliers and a zero effect of $Z$ among non-compliers (whose treatment does not react to the instrument). The denominator consists of the first-stage effect,  i.e.\ the impact of the lottery assignment $Z$ on the decision to reside in Liechtenstein one year after the first lottery participation $D$.
\vspace{-.5cm}
\begin{equation}
\centering
LATE =   \frac{E[Y \cdot Z /\Pr(Z=1|X)- Y \cdot(1-Z)/(1-\Pr(Z=1|X))]}{E[D \cdot Z /\Pr(Z=1|X)- D \cdot(1-Z)/(1-\Pr(Z=1|X)]}\label{LATE}
\end{equation}
For the estimation of \eqref{LATE}, we use the `lateweight' command of the `causalweight' package \citep{Bodory2018} for the statistical software R, with 1999 bootstrap replications for computing the standard error and the default trimming rule of dropping observations with propensity scores smaller than 0.05 or larger than 0.95 to ensure common support in the sample. The instrument propensity score $\Pr(Z=1|X)$ is estimated by means of a probit specification. However, we point out that our estimator is fully nonparametric when controlling for lottery period dummies only, which amounts to a fully saturated model. Our estimator is semiparametric when additionally controlling for further covariates and in particular age, whose inclusion in the linear index of the probit model imposes parametric assumptions on  $\Pr(Z=1|X)$ (but in contrast to two-stage least squares neither on the treatment, nor on the outcome model).

\section{Results}\label{s:results}

This section provides the empirical results. First, the average LATE estimates when pooling all outcome periods; second, the outcome period-specific LATE estimates; and third, an analysis of effect heterogeneity.\footnote{We also briefly discuss the results when considering the second and third (rather than the first) lottery participation as instrument and present the results of these further analyses in Appendix~\ref{s:app_robustness}.} 
 Pooling the outcome periods provides a weighted average of effects over different complier groups and outcome periods, in which compliers who first participate in the lottery in an earlier period obtain a larger weight due to having a longer outcome window than compliers participating in a later period. Furthermore, earlier outcome periods obtain a larger weight than later ones, as earlier outcome periods (e.g.\ two years after first lottery participation, $t=2$) are also observed for first lottery participants in later periods, while the observability of later outcome periods (e.g.\ ten years after first lottery participation, $t=10$) is conditional on a relatively early participation in the lottery. While pooling and its implied weighting of observations might be considered as hampering the interpretability of the results, our outcome period-specific results presented further below suggest that the LATEs on the binary employment and residence decision as well as the activity level (in \%) are quite persistent across different choices of $t$. Given that the effects are quite stable across periods, pooling yields a concise and informative LATE and at the same time entails a higher statistical power (or a smaller standard error) than outcome period-specific estimations that rely on a relatively small subsample of the data.

Table \ref{tab:Empirical results: First participation only year dummies} reports the LATE estimates for pooled outcome periods. The upper panel reports the effects along with bootstrap-based standard error and p-values obtained from t-tests.  As an individual might be observed in multiple outcome periods, we cluster observations on the personal identifier by using the cluster or block bootstrap (which resamples individuals with all related observations in any outcome period rather than single observations) when computing standard errors. We find that having moved to Liechtenstein one year after the first lottery participation increases the probability of residing in Liechtenstein by 71 percentage points and the probability of being employed in Liechtenstein by 24 percentage points among compliers when averaging over all outcome periods. Similarly, the effect on the activity level, which is measured in percent and by definition zero if not working in Liechtenstein, amounts to almost 20 percentage points. Furthermore, the duration of residing and being employed in Liechtenstein increases by 3.44 and 1.15 years on average, respectively, across the outcome periods, which start with $t=2$ and are restricted by the time window of the data set. These important labor market and residential effects are highly statistically significant, as p-values are close to zero. The intermediate panel reports the first-stage effect of  the instrument on the treatment, which implies that 36\% of the participants in the sample are compliers. The group of non-compliers also contains participants who won the pre-draw of the lottery but not the final draw, as we use the pre-draw  as our instrument.  The p-value of the first-stage is close to zero and the instrument is therefore strongly associated with the treatment, thus supporting the relevance assumption postulated in \eqref{relevance}. For completeness, the lower panel of Table \ref{tab:Empirical results: First participation only year dummies} reports the intention-to-treat effect (ITT) of the instrument on the outcome, which is smaller than the corresponding LATE due to the presence of non-compliers for whom the effect is zero by definition (if defiers do not exist). All effects are highly statistically significant. As no extremely high ($>$0.95) or extremely low ($<$0.05) probabilities of winning the lottery occur in any year of first lottery participation, no observation was trimmed such that the estimates are based on all 20,009 pooled observations, as indicated at the bottom of Table \ref{tab:Empirical results: First participation only year dummies}.

Table \ref{tab:Empirical results: First participation with covariates} in Appendix~\ref{s:app_robustness} provides the results for pooled outcome periods when including age, gender, nationality, and missing dummies for these variables as covariates in addition to the lottery period dummies. The effect estimates are rather similar and again highly statistically significant. Furthermore, Tables~\ref{tab:Empirical results: Second participation only year dummies} to \ref{tab:Empirical results: Third participation with covariates} in Appendix~\ref{s:app_robustness} report the estimates for pooled outcome periods when considering the second and third lottery participation (rather than the first one) as instrument, respectively, when either using the lottery period dummies alone or additionally the personal characteristics as control variables. Also in these cases, the findings are all qualitatively similar to our main results.


\begin{table}[!h]
\footnotesize
\center
\caption{Empirical results based on first participation and year dummies}
\label{tab:Empirical results: First participation only year dummies}
\begin{center}
{
\begin{tabular}{l|c|c|c|c|c}
 \hline\hline
   & \multicolumn{5}{c}{Outcomes} \\
  \hline
& Residing (binary) & Employed (binary)  & Activity level (\%)  & Years residing  & Years employed   \\
  \hline
 \multicolumn{6}{c}{LATE} \\
 \hline
Effect & 0.71 & 0.24 & 19.72 & 3.44 & 1.15 \\
Standard error & 0.05 & 0.08 & 7.43 & 0.27 & 0.39 \\
 P-value & 0.00 & 0.00 & 0.01 & 0.00 & 0.00 \\
 \hline
 \multicolumn{6}{c}{First-stage} \\
 \hline
Effect & \multicolumn{1}{c}{}& \multicolumn{1}{c}{}& \multicolumn{1}{c}{0.36 } &\multicolumn{1}{c}{}&\\
 Standard error & \multicolumn{1}{c}{}& \multicolumn{1}{c}{}& \multicolumn{1}{c}{ 0.03 } &\multicolumn{1}{c}{}&\\
 P-value & \multicolumn{1}{c}{}& \multicolumn{1}{c}{}& \multicolumn{1}{c}{0.00} &\multicolumn{1}{c}{}&\\
  \hline
  \multicolumn{6}{c}{ITT} \\
  \hline
Effect & 0.25 & 0.09 & 7.06 & 1.23 & 0.41 \\
Standard error & 0.03 & 0.03 & 2.92 & 0.15 & 0.15 \\
 P-value & 0.00 & 0.00 & 0.02 & 0.00 & 0.01 \\
 \hline
   Number of observations & \multicolumn{1}{c}{}& \multicolumn{1}{c}{}& \multicolumn{1}{c}{20,009} &\multicolumn{1}{c}{}&\\
  Trimmed observations & \multicolumn{1}{c}{}& \multicolumn{1}{c}{}&  \multicolumn{1}{c}{0} &\multicolumn{1}{c}{}&\\
  \hline
\end{tabular}

}
\end{center}

    \par
{\footnotesize Note: Standard errors are estimated by cluster bootstrapping.}
\end{table}

\begin{figure}[h!]
     \begin{center}
        \subfigure[Residing (binary)]{%
            \label{fig:first}
      \includegraphics[width=0.3\textwidth]{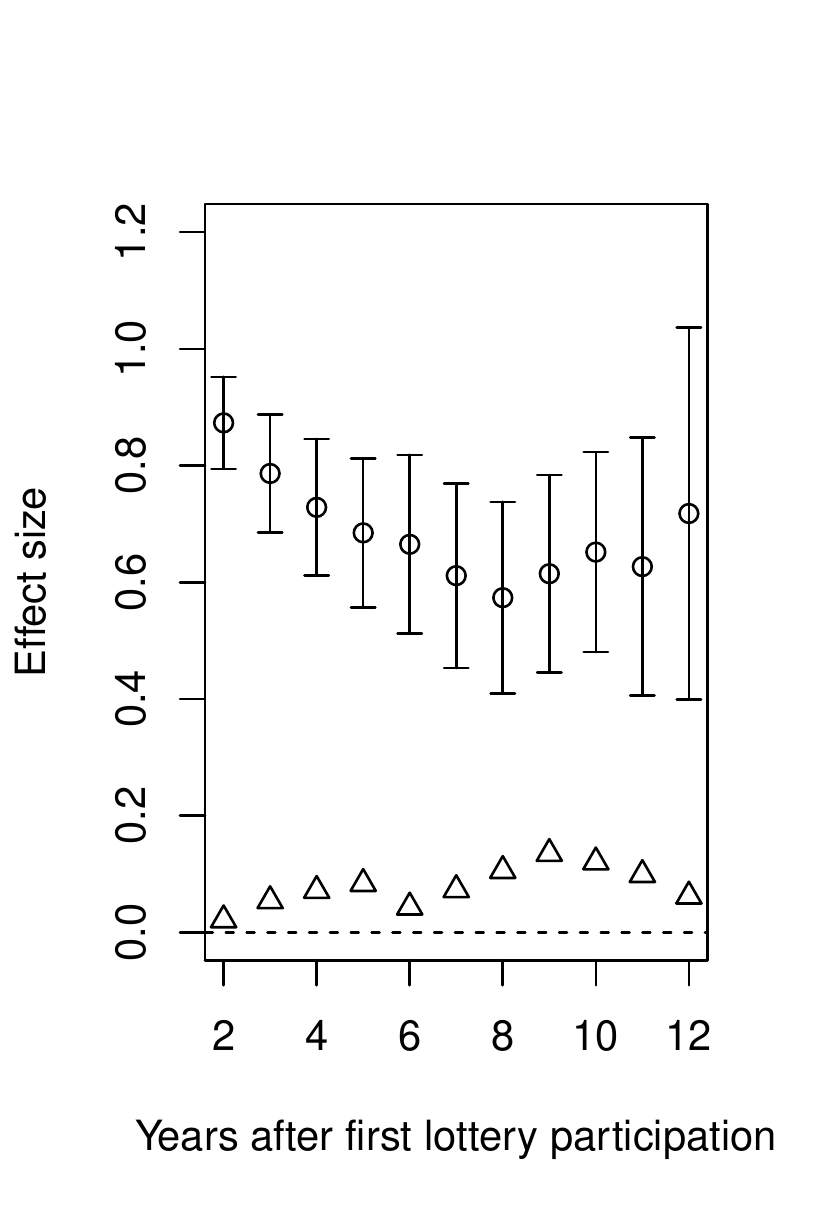}
       }%
     \subfigure[Employed (binary)]{%
            \label{fig:second}
            \includegraphics[width=0.3\textwidth]{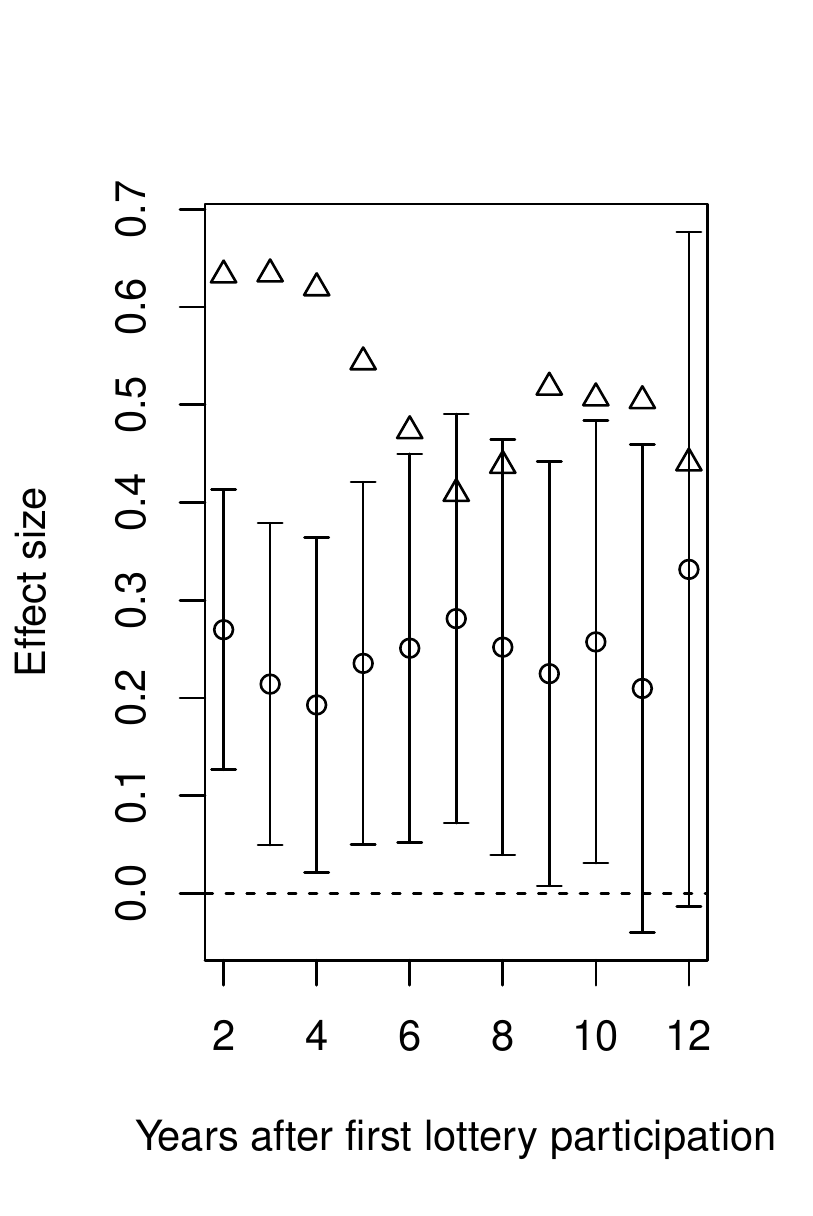}
        }%
        \subfigure[Activity level (\%)]{%
            \label{fig:third}
            \includegraphics[width=0.3\textwidth]{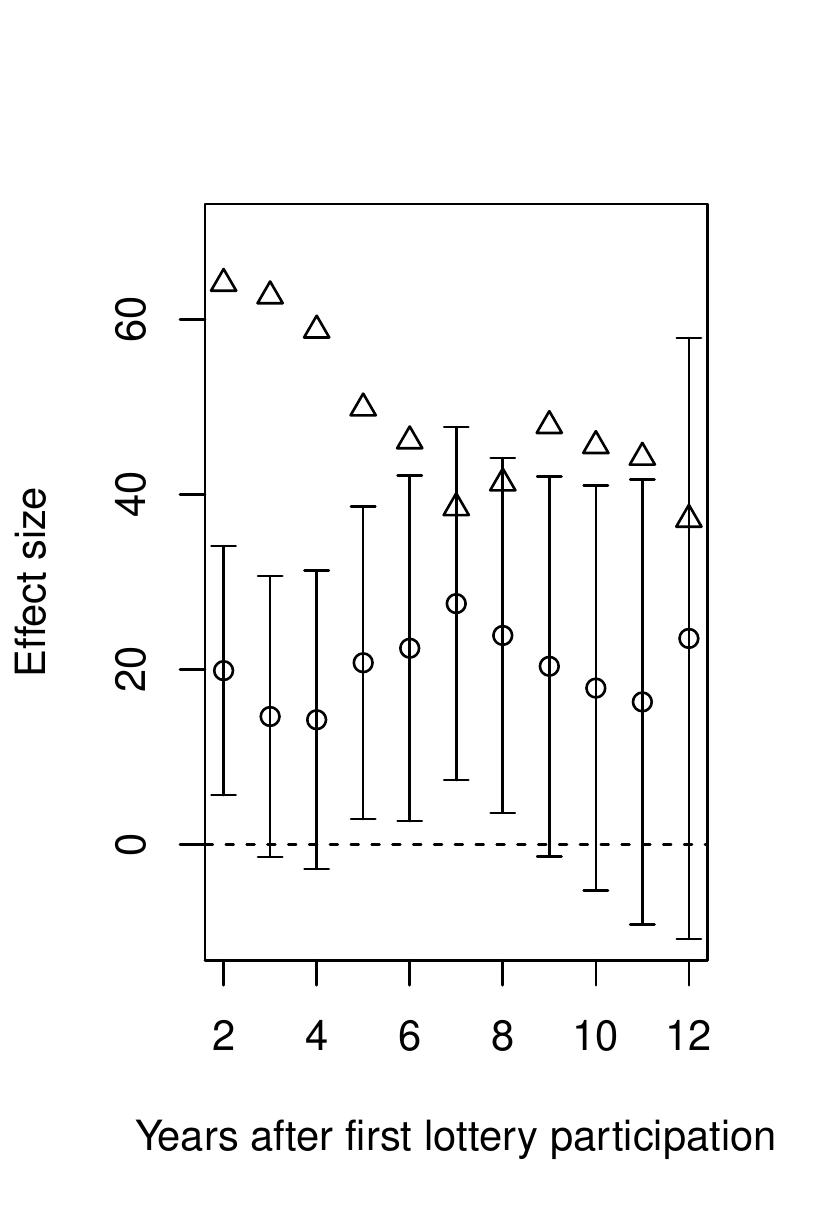}
    }\\ 
        \subfigure[Years residing]{%
            \label{fig:first}
      \includegraphics[width=0.3\textwidth]{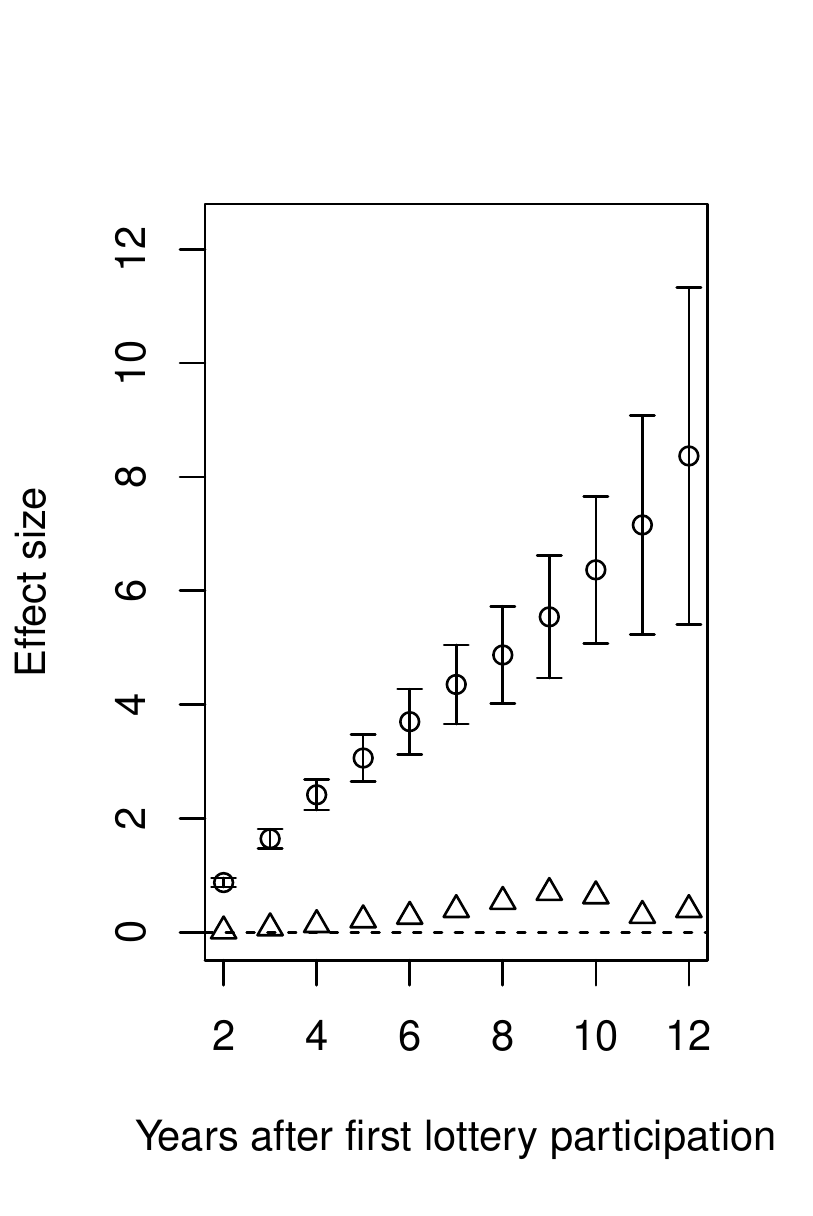}
       }%
        \subfigure[Years employed]{%
           \label{fig:second}
  \includegraphics[width=0.3\textwidth]{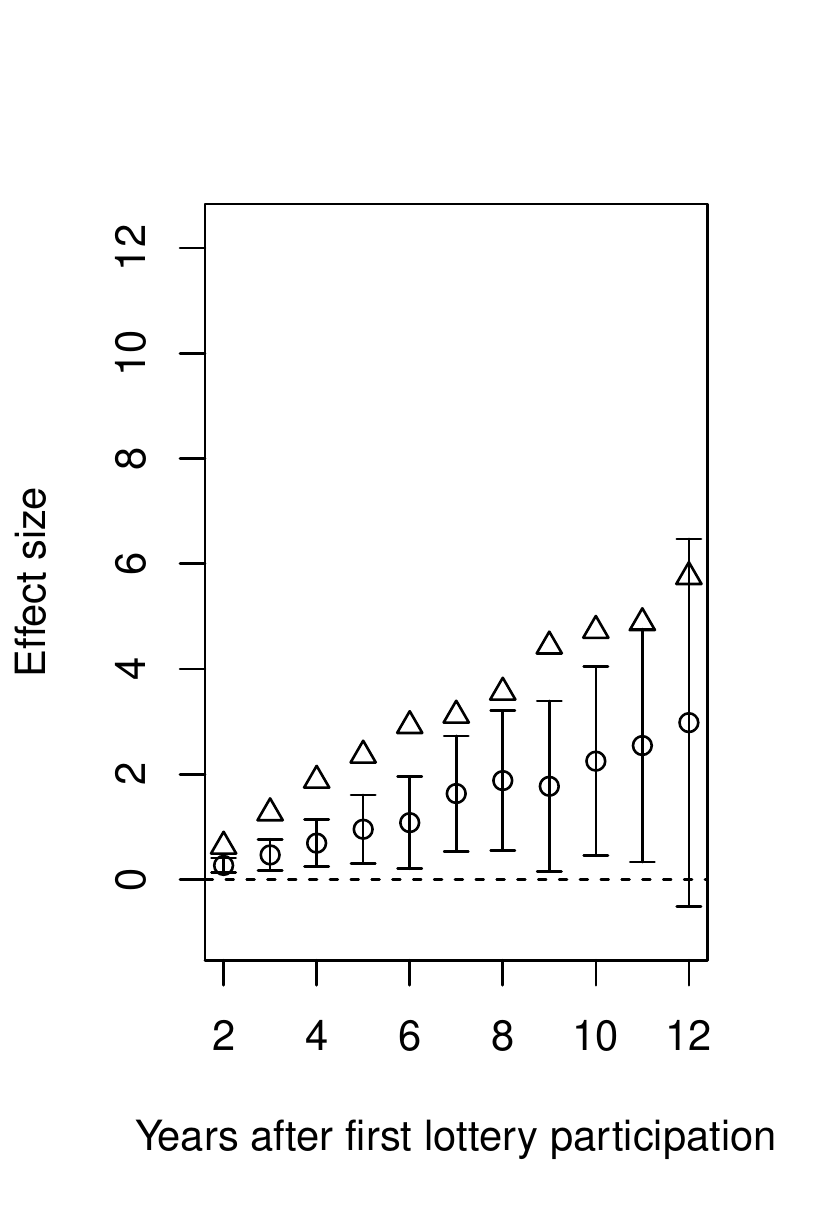}
                }%
        \subfigure[Number of observations]{%
            \label{fig:third}
            \scalebox{0.85}{
  \includegraphics[width=0.3\textwidth]{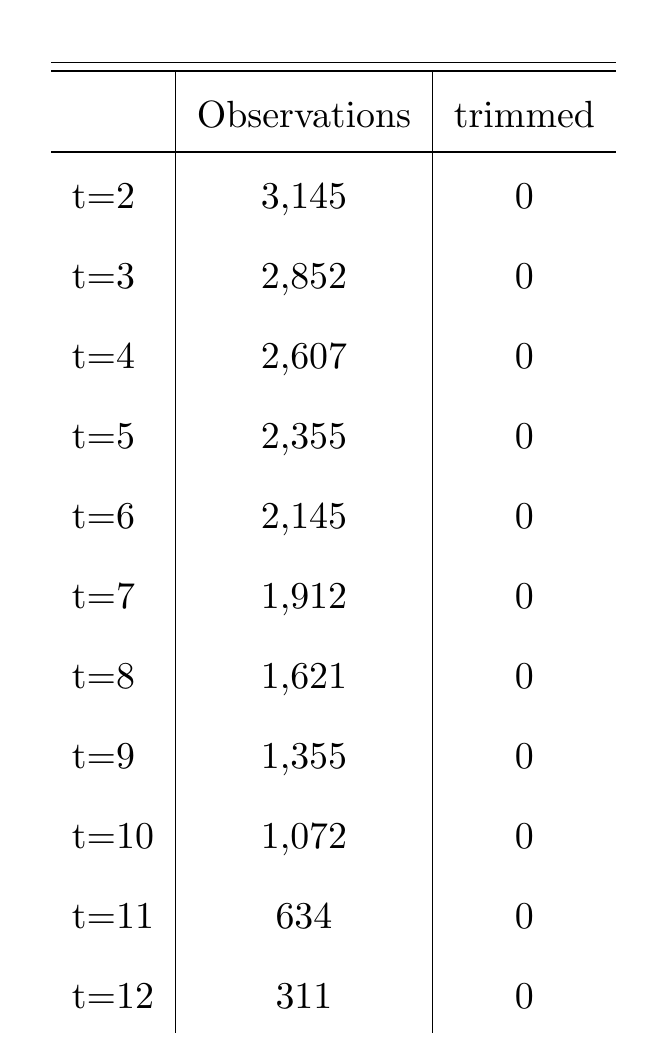}
}
}

  \end{center}
    \caption{%
        Effects over years. Dots represent LATEs, bands correspond to 95\% confidence intervals, triangles depict mean potential outcome among compliers under non-treatment  \label{fig:over_years}
     }%
\end{figure}

In a next step, we investigate the effects in specific outcome periods defined relative to the year of the first lottery participation. Figure~\ref{fig:over_years} displays the estimates for the various outcomes from period $t=2$ (i.e.\ two years after the lottery) up to period $t=12$. The dots represent the period-specific LATEs and the bands correspond to the pointwise 95\% confidence intervals based on the standard bootstrap. \bb{The effects are significantly different from zero.} The triangles depict the estimated mean potential outcome among compliers under-nontreatment, see for instance \cite{Huber2019} for a discussion of its computation, in order to judge the importance of the effects relative to not moving to Liechtenstein.
The effects on the binary residence and employment dummies as well as the activity level are positive throughout all periods and statistically significant at the 5\% level in most cases. However,  many of the effects are imprecisely estimated in particular in later outcome periods with a the limited number of observations, which results in wider confidence intervals. Nevertheless, the positive point estimates appear to be quite persistent with no sign of fading out at the end of the data window. Relatedly, the LATEs on the durations of residing or being employed in Liechtenstein from $t=2$ on monotonically increase as we consider later outcome periods and due to the persistence of the residence and labor market decisions over time, they appear to roughly follow a linear path. Panel~(f) of Figure~\ref{fig:over_years} provides the number of observations per outcome period as well as the number of trimmed observations, which is equal to zero just as for the pooled estimations.  We also inspected the plots when controlling  for age, gender, nationality, and missing dummies in addition to the lottery period dummies and obtained similar results.

	\clearpage
	
	
Approximately half of the observations in our evaluation data consist of participants that had already worked as cross-border commuters one year prior to their first lottery, the other half of potentially new labor market entrants were not employed in and thus, not commuting to Liechtenstein in the previous year. 
Tables \ref{descr_non-cross-border} and \ref{descr_cross-border} report  descriptive statistics about personal characteristics separately for  cross-border commuters and non-commuters in our evaluation sample. We see from the tables that the majority of  applicants is  male, on average 36 to 38 years old, and  of either Austrian or German nationality. However, the share of the nationalities differs in both groups: German is  the most frequent nationality  among the non-commuters (on average 47\%), whereas  the Austrian nationality  dominates  among the cross-border commuters (on average 44\%).\footnote{As mentioned in Section~\ref{s:data}, nationality  is missing for some non-commuters, due to missing information in the lottery data that could not be compensated by information in the employment statistics.}

\begin{table}[!h]
\center
\footnotesize
\caption{Descriptive statistics for non-commuters: First participation from 2006 to 2016}
\label{descr_non-cross-border}
\begin{center}
 \resizebox{\linewidth}{!}{
\begin{tabular}{lcc|cc|ccc|c}
 \hline\hline
   & \multicolumn{2}{c|}{$Z=1$} & \multicolumn{2}{c|}{$Z=0$} & & & \\
  \hline
 & mean & std.dev & mean & std.dev & mean difference & t-value & p-value & 
  observations \\
  \hline
Female & 0.29 & 0.46 & 0.31 & 0.46 & -0.02 & -0.43 & 0.67 & 1,530 \\
[0.15cm]
    \textit{Nationality}  & & & & & & &\\
  Missing Dummy  & 0.00 & 0.00 & 0.03 & 0.18 & -0.03 & -6.82 & 0.00 & 1,530\\
  Austria & 0.29 & 0.45 & 0.29 & 0.45 & 0.00 & 0.03 & 0.97 & 1,485 \\
  Germany & 0.43 & 0.50 & 0.48 & 0.50 & -0.05 & -1.21 & 0.23 & 1,485 \\
  Italy & 0.08 & 0.27 & 0.07 & 0.25 & 0.01 & 0.35 & 0.72 & 1,485 \\
  Switzerland & 0.00 & 0.00 & 0.01 & 0.08 & -0.01 & -2.83 & 0.00 & 1,485 \\
  Others & 0.20 & 0.40 & 0.16 & 0.36 & 0.05 & 1.42 & 0.16 & 1,485 \\
  [0.15cm]
    \textit{Age}  & & & & & &  &\\
  Missing Dummy  & 0.02 & 0.14 & 0.05 & 0.21 & -0.03 & -2.22 & 0.03 & 1,530 \\
    Age & 37.06 & 9.46 & 38.74 & 10.12 & -1.67 & -2.06 & 0.04 & 1,463 \\[0.15cm]

\textit{First lottery participation}  & & & & & & & \\
  Dummy 2006 & 0.11 & 0.31 & 0.11 & 0.32 & -0.00 & -0.17 & 0.86 & 1,530 \\
  Dummy 2007 & 0.11 & 0.32 & 0.11 & 0.32 & 0.00 & 0.04 & 0.97 & 1,530 \\
  Dummy 2008 & 0.09 & 0.29 & 0.16 & 0.36 & -0.07 & -2.71 & 0.01 & 1,530 \\
  Dummy 2009 & 0.10 & 0.30 & 0.09 & 0.29 & 0.01 & 0.45 & 0.65 & 1,530 \\
  Dummy 2010 & 0.06 & 0.24 & 0.09 & 0.28 & -0.02 & -1.02 & 0.31 & 1,530 \\
   Dummy 2011 & 0.12 & 0.33 & 0.09 & 0.28 & 0.03 & 1.18 & 0.24 & 1,530 \\
   Dummy 2012 & 0.10 & 0.29 & 0.07 & 0.25 & 0.03 & 1.13 & 0.26 & 1,530 \\
   Dummy 2013 & 0.07 & 0.26 & 0.05 & 0.22 & 0.02 & 1.00 & 0.32 & 1,530 \\
   Dummy 2014 & 0.07 & 0.26 & 0.06 & 0.24 & 0.01 & 0.31 & 0.76 & 1,530 \\
   Dummy 2015 & 0.10 & 0.29 & 0.08 & 0.27 & 0.02 & 0.65 & 0.51 & 1,530 \\
   Dummy 2016 & 0.07 & 0.26 & 0.09 & 0.29 & -0.02 & -1.06 & 0.29 & 1,530 \\
    \hline
   Number of observations& 157 &  & 1,373 &  &  &  &  &  \\
     \hline
\end{tabular}
}\\
 {\footnotesize Sources: Lottery data (2005 - 2016) and employment statistics (2005 - 2016).\\
  \par}
\end{center}
\par

\end{table}

\begin{table}[!h]
\center
\footnotesize
\caption{Descriptive statistics for cross-border commuters: First participation from 2006 to 2016}
\label{descr_cross-border}
\begin{center}
 \resizebox{\linewidth}{!}{
\begin{tabular}{lcc|cc|ccc|c}
 \hline\hline
   & \multicolumn{2}{c|}{$Z=1$} & \multicolumn{2}{c|}{$Z=0$} & & & \\
  \hline
 & mean & std.dev & mean & std.dev & mean difference & t-value & p-value & 
  observations \\
  \hline
Female & 0.28 & 0.45 & 0.29 & 0.45 & -0.01 & -0.25 & 0.81 & 1,615 \\
[0.15cm]
    \textit{Nationality}  & & & & & & &\\
  Austria & 0.45 & 0.50 & 0.44 & 0.50 & 0.01 & 0.24 & 0.81 & 1,615 \\
  Germany & 0.36 & 0.48 & 0.36 & 0.48 & 0.00 & 0.13 & 0.90 & 1,615 \\
  Italy & 0.05 & 0.21 & 0.08 & 0.26 & -0.03 & -1.70 & 0.09 & 1,615 \\
  Switzerland & 0.02 & 0.12 & 0.01 & 0.07 & 0.01 & 1.08 & 0.28 & 1,615 \\
  Others & 0.12 & 0.33 & 0.12 & 0.32 & 0.00 & 0.19 & 0.85 & 1,615 \\
[0.15cm]
  Age & 37.40 & 9.11 & 36.35 & 8.99 & 1.05 & 1.50 & 0.13 & 1,615 \\
[0.15cm]
\textit{First lottery participation}  & & & & & & & \\
  Dummy 2006 & 0.08 & 0.28 & 0.09 & 0.28 & -0.00 & -0.17 & 0.87 & 1,615 \\
  Dummy 2007 & 0.08 & 0.27 & 0.09 & 0.29 & -0.02 & -0.79 & 0.43 & 1,615 \\
  Dummy 2008 & 0.10 & 0.30 & 0.13 & 0.34 & -0.04 & -1.50 & 0.13 & 1,615 \\
 Dummy 2009 & 0.10 & 0.31 & 0.09 & 0.28 & 0.02 & 0.74 & 0.46 & 1,615 \\
  Dummy 2010 & 0.06 & 0.24 & 0.09 & 0.29 & -0.03 & -1.43 & 0.15 & 1,615 \\
 Dummy 2011 & 0.09 & 0.29 & 0.09 & 0.29 & 0.00 & 0.02 & 0.98 & 1,615\\
  Dummy 2012 & 0.08 & 0.28 & 0.08 & 0.27 & 0.01 & 0.30 & 0.77 & 1,615 \\
  Dummy 2013 & 0.09 & 0.29 & 0.08 & 0.27 & 0.01 & 0.59 & 0.56 & 1,615 \\
  Dummy 2014 & 0.13 & 0.34 & 0.09 & 0.29 & 0.04 & 1.52 & 0.13 & 1,615 \\
  Dummy 2015 & 0.08 & 0.28 & 0.07 & 0.26 & 0.01 & 0.43 & 0.67 & 1,615 \\
  Dummy 2016 & 0.09 & 0.29 & 0.10 & 0.29 & -0.00 & -0.11 & 0.92 & 1,615 \\
   \hline
   Number of observations& 193&  & 1,422 &  &  &  &  &  \\
    \hline
\end{tabular}
}\\
 {\footnotesize Sources: Lottery data (2005 - 2016) and employment statistics (2005 - 2016).\\
  \par}
\end{center}
\par

\end{table}

In a next step, we check whether cross-border commuters who apply for the lottery are in terms of their personal characteristics similar to or different from cross-border commuters to Liechtenstein in general. For this reason, we compare the average age, gender, and  nationality of the cross-border commuters in our sample with those respective average values in the administrative statistics \citep{Beschaftigungsstatistik2018,Beschaftigungsstatistik2019}. Male (71\% versus 74\%) and younger (37 years versus 41 years) occur more frequently in our lottery data. While Austrian and Germans cross-border commuters are the most frequent applicants in our data,
 they are still ``underrepresented,'' as 55\% of  the cross-border commuters with EEA nationality are Austrian and 24\% German. Intuitively, Austrians and Germans who can commute from their home country are less likely to apply for the lottery than cross-border commuters with other nationalities.
   Additionally, we use the cross-border survey to draw conclusions about what the average educational level might be in our evaluation data, in which information on education is not available. \cite{Marxer2016} report that cross-border commuters  are on average highly educated, as 57.5\% of them hold a degree  from a higher education institution. Based on these findings, we suspect that cross-border commuters applying for the migration lottery have an average education that is likely considerably higher than that of the general population in Austria, Switzerland, and Liechtenstein.

From a policy perspective, it appears interesting whether treatment effects are heterogeneous across cross-border commuters and non-commuters, i.e.\ if residence permits are rather effective for attracting new or keeping existing foreign workers that have already decided to enter Liechtenstein's labor market at an earlier point in time. If permits were more effective among one rather than the other group, policy makers might want to consider to adapt targeting of immigration policies accordingly.


\begin{table}[!h]
\footnotesize
\center
\caption{Effects among non-commuters}
\label{tab:Heterogenous effects: Non-cross-border commuters}
\begin{center}
{
\begin{tabular}{l|c|c|c|c|c}
 \hline\hline
   & \multicolumn{5}{c}{Outcomes} \\
  \hline
& Residing (binary) & Employed (binary)  & Activity level (\%)  & Years residing  & Years employed   \\
  \hline
 \multicolumn{6}{c}{LATE} \\
 \hline
Effect & 0.71 & 0.34 & 29.78 & 3.31 & 1.56 \\
Standard error  & 0.13 & 0.15 & 13.83 & 0.60 & 0.79 \\
P-value  & 0.00 & 0.02 & 0.03 & 0.00 & 0.05 \\
\hline
 \multicolumn{6}{c}{First-stage} \\
 \hline
Effect & \multicolumn{1}{c}{}& \multicolumn{1}{c}{}& \multicolumn{1}{c}{0.28 } &\multicolumn{1}{c}{}&\\
 Standard error & \multicolumn{1}{c}{}& \multicolumn{1}{c}{}& \multicolumn{1}{c}{ 0.05 } &\multicolumn{1}{c}{}&\\
 P-value & \multicolumn{1}{c}{}& \multicolumn{1}{c}{}& \multicolumn{1}{c}{0.00}&\multicolumn{1}{c}{}&\\
  \hline
  \multicolumn{6}{c}{ITT} \\
  \hline
Effect & 0.20 & 0.10 & 8.35 & 0.93 & 0.44 \\
Standard error  & 0.04 & 0.04 & 4.13 & 0.21 & 0.23 \\
P-value  & 0.00 & 0.03 & 0.04 & 0.00 & 0.06 \\
  \hline
   Number of observations & \multicolumn{1}{c}{}& \multicolumn{1}{c}{}& \multicolumn{1}{c}{10,081} &\multicolumn{1}{c}{}&\\
 Trimmed observations & \multicolumn{1}{c}{}& \multicolumn{1}{c}{}&  \multicolumn{1}{c}{0} &\multicolumn{1}{c}{}&\\
  \hline
\end{tabular}

}
\end{center}

    \par
{\footnotesize Note: Standard errors are estimated by cluster bootstrapping.}
\end{table}

For this reason, Tables \ref{tab:Heterogenous effects: Non-cross-border commuters} and \ref{tab:Heterogenous effects: Cross-border commuters} report the results with pooled outcome periods separately for applicants working (cross-border commuters) and not working (non-commuters) in Liechtenstein one year prior to the lottery. We find that in both subsamples, the residence permit has a similarly positive and highly significant effect on the compliers' probability to reside in Liechtenstein (71 vs.\ 69 percentage points) and their residence duration (3.31 vs.\ 3.46 years). In contrast, we find heterogeneous effects for the LATEs on the labor market outcomes: The effects  for previous cross-border commuters are insignificant, whereas the effects  for people not working in Liechtenstein  are   significant at conventional levels. For the latter group of potential new labor market entrants, we find that a residence permit leads to an increase in the employment probability of 34\%, in the activity level of almost 30\%, and in the employment duration of 1.56 years in the outcome periods. We therefore conclude that the policy is more effective in raising labor supply among individuals previously not working in Liechtenstein than among cross-border commuters, while effects on residential choices are similar among both groups.
\clearpage
\begin{table}[!h]
\footnotesize
\center
\caption{Effects among cross-border commuters}
\label{tab:Heterogenous effects: Cross-border commuters}
\begin{center}
{
\begin{tabular}{l|c|c|c|c|c}
 \hline\hline
   & \multicolumn{5}{c}{Outcomes} \\
  \hline
& Residing (binary) & Employed (binary)  & Activity level (\%)  & Years residing  & Years employed   \\
  \hline
 \multicolumn{6}{c}{LATE} \\
 \hline
Effect & 0.69 & 0.11 & 6.75 & 3.46 & 0.56 \\
 Standard error & 0.06 & 0.09 & 8.83 & 0.28 & 0.42 \\
 P-value & 0.00 & 0.21 & 0.44 & 0.00 & 0.18 \\
   \hline
 \multicolumn{6}{c}{First-stage} \\
 \hline
Effect & \multicolumn{1}{c}{}& \multicolumn{1}{c}{}& \multicolumn{1}{c}{0.42 } &\multicolumn{1}{c}{}& \\
 Standard error & \multicolumn{1}{c}{}& \multicolumn{1}{c}{}& \multicolumn{1}{c}{ 0.04 }  &\multicolumn{1}{c}{}&\\
 P-value & \multicolumn{1}{c}{}& \multicolumn{1}{c}{}& \multicolumn{1}{c}{0.00} &\multicolumn{1}{c}{}&\\
  \hline
  \multicolumn{6}{c}{ITT} \\
  \hline

Effect & 0.29 & 0.05 & 2.86 & 1.47 & 0.24 \\
 Standard error & 0.04 & 0.04 & 3.75 & 0.20 & 0.18 \\
 P-value & 0.00 & 0.23 & 0.45 & 0.00 & 0.20 \\
  \hline
  Number of observations & \multicolumn{1}{c}{}& \multicolumn{1}{c}{}& \multicolumn{1}{c}{9,928} &\multicolumn{1}{c}{}&\\
  Trimmed observations & \multicolumn{1}{c}{}& \multicolumn{1}{c}{}& \multicolumn{1}{c}{0} &\multicolumn{1}{c}{}&\\
  \hline
  \end{tabular}

}
\end{center}

    \par
{\footnotesize Note: Standard errors are estimated by cluster bootstrapping.}
\end{table}

\section{Conclusion}\label{s:concl}

In this paper, we analyzed the effect of a residence permit on the labor supply and residential decisions of foreign workers by an instrumental variable approach exploiting a migration lottery in Liechtenstein. Our results pointed to substantial effects on the labor market and residential attachment of compliers, whose migration decision complies with the permit assignment in their first lottery. We also found the labor market effects to be more strongly driven by individuals previously not working in Liechtenstein than by previous cross-border commuters. This implies that residence permits are more effective for attracting new labor market entrants than for keeping previously commuting workers in the labor market and therefore suggests stronger effects at the extensive rather than intensive margin of labor supply. In future research following up on our findings, one would ideally link our current data base with tax data to examine the effect of resident permits on the income (at least) of current cross-border commuters or on the tax revenues in Liechtenstein.






\clearpage

\appendix

\numberwithin{equation}{section}
\numberwithin{lemma}{section}
\numberwithin{proposition}{section}
\numberwithin{figure}{section}
\numberwithin{table}{section}

\section{Appendix}\label{s:app}

\subsection{Detailed institutional background} 
\label{s:app_background} \bb{I have not adapted this appendix to the new section in the main text.} \sg{09.02.2021: I did it.}

This section provides information about the conditions of participation in the lottery  and the draw in more detail.
Lottery participants must hold an EEA citizenship and transfer the required application documents as well as the participation fees prior to a specific deadline \citep[section 38]{PFZG2009}. The amount of the fee varies between the pre-draw (100 CHF) and the final draw (500 CHF) \citep{APA2020}. Persons with an entry ban, posing a threat to public safety, or providing false statements are already excluded from the first draw of the lottery \citep[section 38, 3]{PFZG2009}.

\begin{figure}[h!]
\begin{center}
{ \includegraphics[scale=.5]{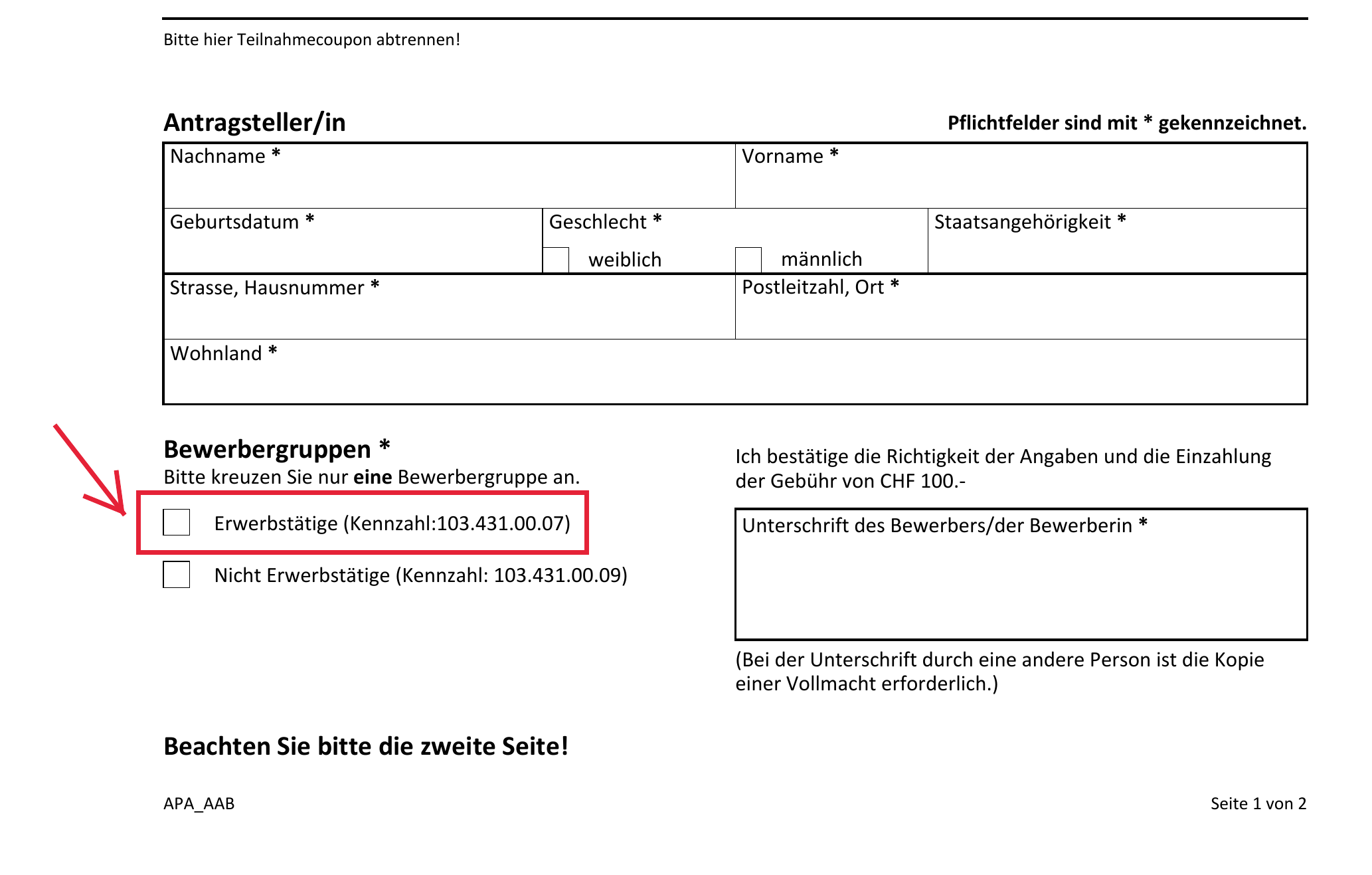}}
 \captionsource{Participation voucher (of 2019). Interested candidates fill out this form to participate in the pre-draw}{\cite{Auslosung19}}
\label{voucher}
 \end{center}
 \end{figure}

\begin{figure}[h!]
\begin{center}
{ \includegraphics[scale=.5]{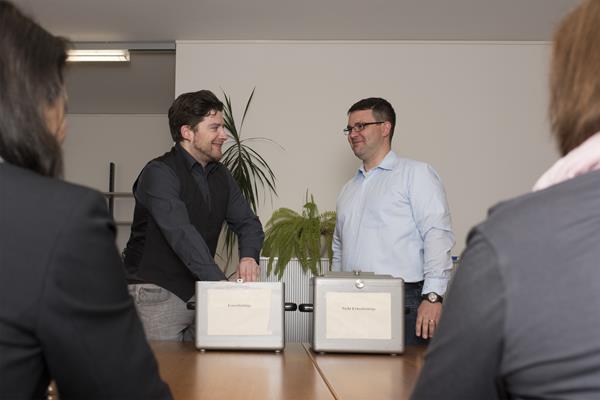}}
 \captionsource{Final draw (of spring lottery 2016) monitored by a judge}{Michael Zanghellini, Liechtensteiner Volksblatt}
 \label{Ziehung}
 \end{center}
 \end{figure}

In the final draw, participants must be of full age and must not hold a permanent residence permit \citep{APA2019}. Importantly, they must also provide an employment contract of more than one year with a minimum activity level of 80\% or an authorized permanent cross-border business activity in case of self-employment \citep{APA2019}.
After winning both the pre-draw and the final draw, the lottery participant must relocate to Liechtenstein within six months, otherwise the residence permit expires \citep[section 37, 2]{PFZG2009}. For this reason, our treatment is defined based on residing in Liechtenstein in the year after the lottery, as obtaining the permit is tied to actually moving there. The drawing procedure can be described as follows. All submitted applications (see Figure \ref{voucher}) are put into a box and even include  participants not fulfilling all conditions (to give them the chance to appeal against a later denial of a residence permit due to a violation of the conditions). In the presence of a national judge and media representatives, the winners are blindly drawn from the box and the person who draws announces the total number of winners as well as their nationality (see Figure~\ref{Ziehung}). Lottery losers may participate again in subsequent lotteries, while multiple applications to the very same lottery are not allowed \citep[section 38, 1) c)]{PFZG2009} . 

\bigskip

\newpage
\clearpage

\subsection{Further analyses and robustness checks}\label{s:app_robustness}

\begin{table}[!h]
\footnotesize
\center
\caption{Empirical results based on first participation and further covariates}
\label{tab:Empirical results: First participation with covariates}
\begin{center}
{
\begin{tabular}{l|c|c|c|c|c }
 \hline\hline
   & \multicolumn{5}{c}{Outcomes} \\
  \hline
& Residing (binary) & Employed (binary)  & Activity level (\%)  & Years residing  & Years employed   \\
  \hline
 \multicolumn{6}{c}{LATE} \\
 \hline
   Effect & 0.70 & 0.21  & 16.70 & 3.43 & 1.02  \\
  Standard error & 0.06 & 0.08  & 7.54 & 0.27  & 0.39 \\
  P-value & 0.00 & 0.01  & 0.03 & 0.00  & 0.01 \\
  \hline
 \multicolumn{6}{c}{First-stage} \\
 \hline
 Effect & \multicolumn{1}{c}{}& \multicolumn{1}{c}{}& \multicolumn{1}{c}{0.35 } &\multicolumn{1}{c}{}&\\
 Standard error & \multicolumn{1}{c}{}& \multicolumn{1}{c}{}& \multicolumn{1}{c}{ 0.03 } &\multicolumn{1}{c}{}&\\
 P-value & \multicolumn{1}{c}{}& \multicolumn{1}{c}{}& \multicolumn{1}{c}{0.00}&\multicolumn{1}{c}{}&\\
  \hline
  \multicolumn{6}{c}{ITT} \\
  \hline

Effect & 0.25 & 0.07  & 5.88 & 1.21 & 0.36  \\
  Standard error & 0.03 & 0.03 & 2.86 & 0.15 & 0.15 \\
  P-value & 0.00  & 0.01 & 0.04  & 0.00 & 0.02 \\
   \hline
    Number of observations & \multicolumn{1}{c}{}& \multicolumn{1}{c}{}& \multicolumn{1}{c}{20,009} &\multicolumn{1}{c}{}&\\
 Trimmed observations & \multicolumn{1}{c}{}& \multicolumn{1}{c}{}& \multicolumn{1}{c}{392} &\multicolumn{1}{c}{}& \\
  \hline
 \end{tabular}
}\\
\end{center}
\par
{\footnotesize Note: Standard errors are estimated by cluster bootstrapping.\\ Only observations whose first lottery participation was after 2005 are included.}

\end{table}

\clearpage

\begin{figure}[h!]
\begin{center}
{ \includegraphics[scale=0.5]{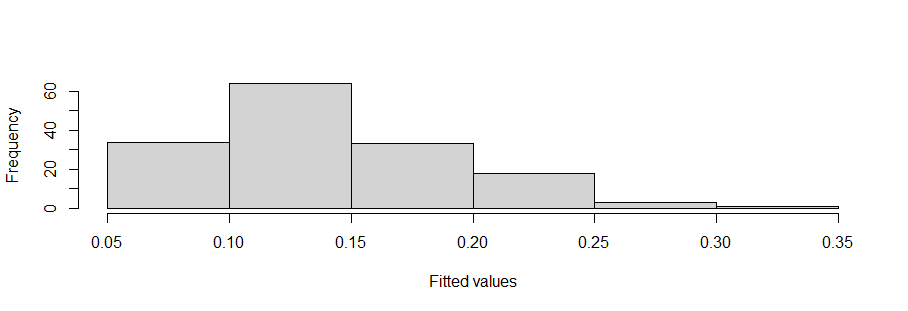}}
\caption{Second participation; Propensity score; Assignment=1}
 \end{center}
 \end{figure}

 \begin{figure}[h!]
\begin{center}
{ \includegraphics[scale=0.5]{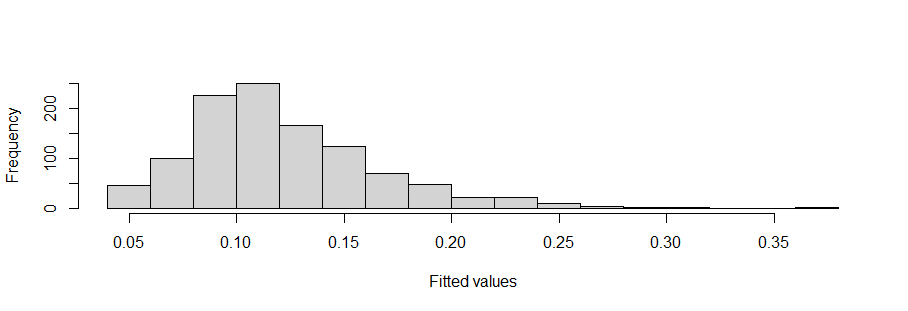}}
\caption{Second participation; Propensity score; Assignment=0}
 \end{center}
 \end{figure}

\clearpage

\begin{table}[!h]
\footnotesize
\center
\caption{Empirical results based on second participation and year dummies}
\label{tab:Empirical results: Second participation only year dummies}
\begin{center}
{
\begin{tabular}{l|c|c|c|c|c}
 \hline\hline
   & \multicolumn{5}{c}{Outcomes} \\
  \hline
& Residing (binary) & Employed (binary)  & Activity level (\%)  & Years residing  & Years employed   \\
  \hline
 \multicolumn{6}{c}{LATE} \\
 \hline
Effect & 0.78 & 0.31 & 27.39 & 3.24 & 1.29 \\
 Standard error & 0.12 & 0.15 & 13.91 & 0.53 & 0.72 \\
  P-value & 0.00 & 0.04 & 0.05 & 0.00 & 0.07 \\

\hline
 \multicolumn{6}{c}{First-stage} \\
 \hline
Effect & \multicolumn{1}{c}{}& \multicolumn{1}{c}{}&\multicolumn{1}{c}{0.40} &\multicolumn{1}{c}{}&\\
 Standard error & 0.06 & 0.07 & 0.07 & 0.07 & 0.07 \\
 P-value & \multicolumn{1}{c}{}& \multicolumn{1}{c}{}&\multicolumn{1}{c}{0.00} &\multicolumn{1}{c}{}&\\
  \hline
  \multicolumn{6}{c}{ITT} \\
  \hline
Effect & 0.31 & 0.12 & 10.84 & 1.28 & 0.51 \\
 Standard error & 0.06 & 0.06 & 5.13 & 0.29 & 0.26 \\
 P-value & 0.00 & 0.03 & 0.03 & 0.00 & 0.05 \\
\hline
  Number of observations & \multicolumn{1}{c}{}& \multicolumn{1}{c}{}& \multicolumn{1}{c}{6,771}&\multicolumn{1}{c}{}&\\
 Trimmed observations & \multicolumn{1}{c}{}& \multicolumn{1}{c}{}& \multicolumn{1}{c}{1,251} &\multicolumn{1}{c}{}&\\
  \hline
\end{tabular}

}
\end{center}

    \par
{\footnotesize Note: Standard errors are estimated by cluster bootstrapping.\\ Only observations whose first lottery participation was after 2005 are included.}

\end{table}

\begin{table}[!h]
\footnotesize
\center
\caption{Empirical results based on second participation and further covariates}
\label{tab:Empirical results: Second participation with covariates}
\begin{center}
{
\begin{tabular}{l|c|c|c|c|c}
 \hline\hline
   & \multicolumn{5}{c}{Outcomes} \\
  \hline
& Residing (binary) & Employed (binary)  & Activity level (\%)  & Years residing  & Years employed   \\
  \hline
 \multicolumn{6}{c}{LATE} \\
  \hline
 Effect & 0.70 & 0.25 & 26.16 & 2.81 & 0.84 \\
 Standard error  & 0.11 & 0.14 & 13.58 & 0.47 & 0.66 \\
 P-value & 0.00 & 0.07 & 0.05 & 0.00 & 0.20 \\
  \hline
  \multicolumn{6}{c}{First-stage} \\
 \hline
 Effect & \multicolumn{1}{c}{}& \multicolumn{1}{c}{}& \multicolumn{1}{c}{0.38} &\multicolumn{1}{c}{}&\\
 Standard error &  \multicolumn{1}{c}{}& \multicolumn{1}{c}{}&\multicolumn{1}{c}{0.06}&\multicolumn{1}{c}{} \\
 P-value & \multicolumn{1}{c}{}& \multicolumn{1}{c}{}& \multicolumn{1}{c}{0.00}&\multicolumn{1}{c}{}&\\
  \hline
  \multicolumn{6}{c}{ITT} \\
  \hline
 Effect & 0.26 & 0.09 & 9.82 & 1.06 & 0.32 \\
 Standard error  & 0.06 & 0.05 & 5.02 & 0.25 & 0.25 \\
 P-value & 0.00 & 0.08 & 0.05 & 0.00 & 0.20 \\
 \hline
  Number of observations & \multicolumn{1}{c}{}& \multicolumn{1}{c}{}& \multicolumn{1}{c}{6,771} &\multicolumn{1}{c}{}&\\
 Trimmed observations & \multicolumn{1}{c}{}& \multicolumn{1}{c}{}&  \multicolumn{1}{c}{1,727} &\multicolumn{1}{c}{}&\\
  \hline
\end{tabular}

}
\end{center}

    \par
{\footnotesize Note: Standard errors are estimated by cluster bootstrapping\\ Only observations whose first lottery participation was after 2005 are included.}

\end{table}

\newpage

\begin{figure}[h!]
\begin{center}
{ \includegraphics[scale=0.5]{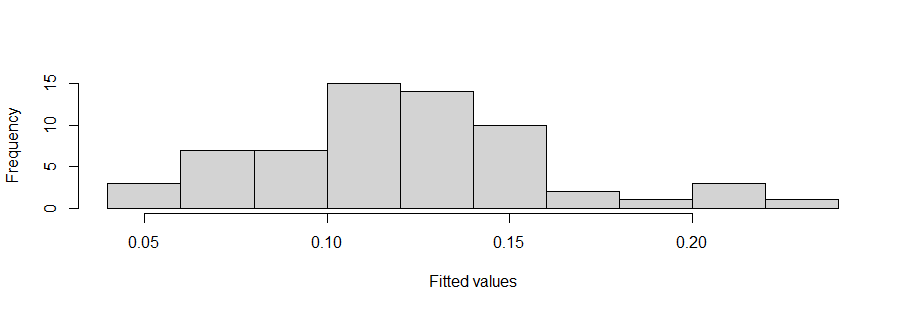}}
\caption{Third participation; Propensity score; Assignment=1}
 \end{center}
 \end{figure}

 \begin{figure}[h!]
\begin{center}
{ \includegraphics[scale=0.5]{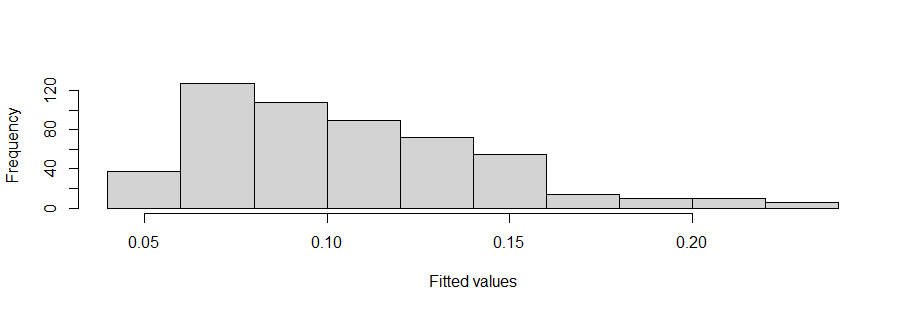}}
\caption{Third participation; Propensity score; Assignment=0}
 \end{center}
 \end{figure}

\clearpage

\begin{table}[!h]
\footnotesize
\center
\caption{Empirical results based on third participation and year dummies }
\label{tab:Empirical results: Third participation only year dummies}
\begin{center}
{
\begin{tabular}{l|c|c|c|c|c}
 \hline\hline
   & \multicolumn{5}{c}{Outcomes} \\
  \hline
& Residing (binary) & Employed (binary)  & Activity level (\%)  & Years residing  & Years employed   \\
  \hline
 \multicolumn{6}{c}{LATE} \\
  \hline
 Effect & 0.77 & 0.22 & 15.19 & 2.99 & 0.88 \\
Standard error  & 0.17 & 0.18 & 16.16 & 0.89 & 0.82 \\
 P-value & 0.00 & 0.22 & 0.35 & 0.00 & 0.28 \\
   \hline
 \multicolumn{6}{c}{First-stage} \\
 \hline
  Effect & \multicolumn{1}{c}{}& \multicolumn{1}{c}{}& \multicolumn{1}{c}{0.61 } &\multicolumn{1}{c}{}&\\
 Standard error & \multicolumn{1}{c}{}& \multicolumn{1}{c}{}& \multicolumn{1}{c}{ 0.10 }&\multicolumn{1}{c}{}& \\
 P-value & \multicolumn{1}{c}{}& \multicolumn{1}{c}{}& \multicolumn{1}{c}{0.00}&\multicolumn{1}{c}{}&\\
  \hline
  \multicolumn{6}{c}{ITT} \\
  \hline
 Effect & 0.47 & 0.13 & 9.21 & 1.81 & 0.53 \\
Standard error  & 0.10 & 0.09 & 8.26 & 0.53 & 0.39 \\
 P-value & 0.00 & 0.16 & 0.27 & 0.00 & 0.17 \\
  \hline
   Number of observations & \multicolumn{1}{c}{}& \multicolumn{1}{c}{}& \multicolumn{1}{c}{3,369} &\multicolumn{1}{c}{}&\\
  Trimmed observations & \multicolumn{1}{c}{}& \multicolumn{1}{c}{}&  \multicolumn{1}{c}{1,088} &\multicolumn{1}{c}{}&\\
  \hline
\end{tabular}

}
\end{center}

    \par
{\footnotesize Note:  Standard errors are estimated by cluster bootstrapping.\\ Only observations whose first lottery participation was after 2005 are included.}

\end{table}

\begin{table}[!h]
\footnotesize
\center
\caption{Empirical results based on third participation and further covariates}
\label{tab:Empirical results: Third participation with covariates}
\begin{center}
{
\begin{tabular}{l|c|c|c|c|c}
 \hline\hline
   & \multicolumn{5}{c}{Outcomes} \\
  \hline
& Residing (binary) & Employed (binary)  & Activity level (\%)  & Years residing & Years employed   \\
  \hline
 \multicolumn{6}{c}{LATE} \\
  \hline
 Effect & 0.80 & 0.24 & 13.87 & 3.58 & 0.98 \\
 Standard error & 0.18 & 0.20 & 18.68 & 0.94 & 0.98 \\
 P-value & 0.00 & 0.22 & 0.46 & 0.00 & 0.32 \\
   \hline
 \multicolumn{6}{c}{First-stage} \\
 \hline
  Effect & \multicolumn{1}{c}{}& \multicolumn{1}{c}{}& \multicolumn{1}{c}{0.50}&\multicolumn{1}{c}{}& \\
 Standard error & 0.10 & 0.11 & 0.10 & 0.11 & 0.11 \\
 P-value & \multicolumn{1}{c}{}& \multicolumn{1}{c}{}& \multicolumn{1}{c}{0.00}&\multicolumn{1}{c}{}&\\
  \hline
  \multicolumn{6}{c}{ITT} \\
  \hline
 Effect & 0.39 & 0.12 & 6.87 & 1.77 & 0.49 \\
 Standard error & 0.09 & 0.10 & 8.94 & 0.50 & 0.45 \\
 P-value& 0.00 & 0.22 & 0.44 & 0.00 & 0.27 \\
    \hline
Number of observations  & \multicolumn{1}{c}{}& \multicolumn{1}{c}{}& \multicolumn{1}{c}{3,369} &\multicolumn{1}{c}{}&\\
Trimmed observations & \multicolumn{1}{c}{}& \multicolumn{1}{c}{}&  \multicolumn{1}{c}{1,193} &\multicolumn{1}{c}{}&\\
  \hline
\end{tabular}

}
\end{center}

    \par
{\footnotesize Note:  Standard errors are estimated by cluster bootstrapping.\\ Only observations whose first lottery participation was after 2005 are included.}

\end{table}

\clearpage

\setlength{\bibsep}{0.7\baselineskip}

\bibliography{LiLoResearch}{}

\end{document}